\documentclass[11pt,preprint]{aastex}

\newcommand{\teff}{${\rm T}_{eff}$}
\newcommand{\ldl}{${\lambda}/{\Delta}{\lambda}$}
\newcommand{\water}{H$_2$O}
\newcommand{\meth}{CH$_4$}

\slugcomment{Accepted to ApJ}

\shorttitle{T Dwarf Classification}
\shortauthors{Burgasser et al.}

\begin{document}

\title{A Unified Near Infrared Spectral Classification Scheme for T Dwarfs}

\author{Adam J.\ Burgasser\altaffilmark{1,2,3},
T.\ R.\ Geballe\altaffilmark{4},
S.\ K.\ Leggett\altaffilmark{5},
J.\ Davy Kirkpatrick\altaffilmark{6}
and
David A.\ Golimowski\altaffilmark{7}}

\altaffiltext{1}{Department of Astrophysics, Division of Physical Sciences,
American Museum of Natural History, Central Park West at 79$^{th}$ Street,
New York, NY 10024; adam@amnh.org}
\altaffiltext{2}{Spitzer Fellow}
\altaffiltext{3}{Now at Massachusetts Institute of Technology,
Kavli Institute for Astrophysics and Space Research, Building 37, 77 Massachusetts Ave,
Cambridge, MA 02139; ajb@mit.edu}
\altaffiltext{4}{Gemini Observatory, 670 North A'ohoku Place, Hilo, HI 96720}
\altaffiltext{5}{Joint Astronomy Centre, 660 North A'ohoku Place, Hilo, HI 96720}
\altaffiltext{6}{Infrared Processing and Analysis Center, M/S
100-22, California Institute of Technology, Pasadena, CA 91125}
\altaffiltext{7}{Department of Physics \& Astronomy, Johns Hopkins University,
3701 San Martin Drive, Baltimore, MD 21218}

\begin{abstract}
A revised near infrared classification scheme for T dwarfs is presented,
based on and superseding prior schemes
developed by Burgasser et al.\ and Geballe et al., and defined
following the precepts of the MK Process. Drawing from two large
spectroscopic libraries of T dwarfs identified largely in the
Sloan Digital Sky Survey and the Two Micron All Sky Survey, nine
primary spectral standards and five alternate standards spanning
spectral types T0 to T8 are identified that match criteria of
spectral character, brightness, absence of a resolved companion
and accessibility from both northern and southern hemispheres. The
classification of T dwarfs is formally made by the direct
comparison of near infrared spectral data of equivalent resolution
to the spectra of these standards. Alternately, we have redefined
five key spectral indices
measuring the strengths of the
major H$_2$O and CH$_4$ bands in the 1--2.5 $\micron$ region that may be used as a proxy
to direct spectral comparison.  Two methods
of determining T spectral type using these indices are outlined and yield equivalent results.
These classifications are also equivalent to those from prior schemes,
implying that no revision of existing spectral type trends is required.
The one-dimensional scheme presented here
provides a first step toward the observational characterization
of the lowest luminosity brown dwarfs currently known.
Future extensions to incorporate spectral variations
arising from differences in photospheric dust content,
gravity and metallicity are briefly discussed.
A compendium of all currently known T dwarfs with updated classifications is presented.
\end{abstract}

\section{Introduction}

Classification is an important first step in all fields of empirical
natural science, from biology (e.g., the taxonomy of species; Linnaeus 1735) to
chemistry (e.g., the periodic table of elements; Mendeleev 1869) to
several subfields of astronomy (e.g., the Hubble sequence of galaxies; Hubble 1936).
The identification and quantification
of similarities and differences in observed phenomena help to clarify
their governing mechanisms, while
providing a standard framework for our
continually evolving theoretical understanding.

In stellar astronomy, spectral classification has been and remains a powerful
tool, providing insight into the physical
characteristics of stars and stellar populations and
enabling the study of Galactic structure (e.g., Morgan, Sharpless \& Osterbrock 1952).
From the first stellar spectral groups designated by Secchi (1866),
the classification of stars has evolved in complexity and breadth,
largely due to advances in technology and the compilation of large
spectral catalogs (e.g., the Henry Draper [HD] catalog, Cannon \& Pickering 1918-1924).
Nevertheless, nearly all existing stellar classification
schemes remain observationally based.
The most successful follow the MK Process \citep{mor43,mor73,kee76,mor78,gar84,cor94},
a method by which stellar classes are defined by
specific standard stars, and all other
stars are classified by the direct comparison of spectra over a designated
wavelength range and resolution.
This method allows spectral classifications to remain independent of
physical interpretations, concepts which can evolve even as
the spectra themselves generally do not.
Examples of MK classification schemes include
the Michigan Catalogue of Spectral Types for the HD stars \citep{hou75,hou78,hou82,hou88,hou99},
automated classifications through neural network techniques
(e.g., von Hippel et al.\ 1994; Bailer-Jones et al.\ 1998) and classification schemes of normal stars
at UV (e.g., Rountree \& Sonneborn 1993) and near infrared (e.g., Wallace \& Hinkle 1997)
wavelengths.

\begin{figure}
\epsscale{1.0}
\plotone{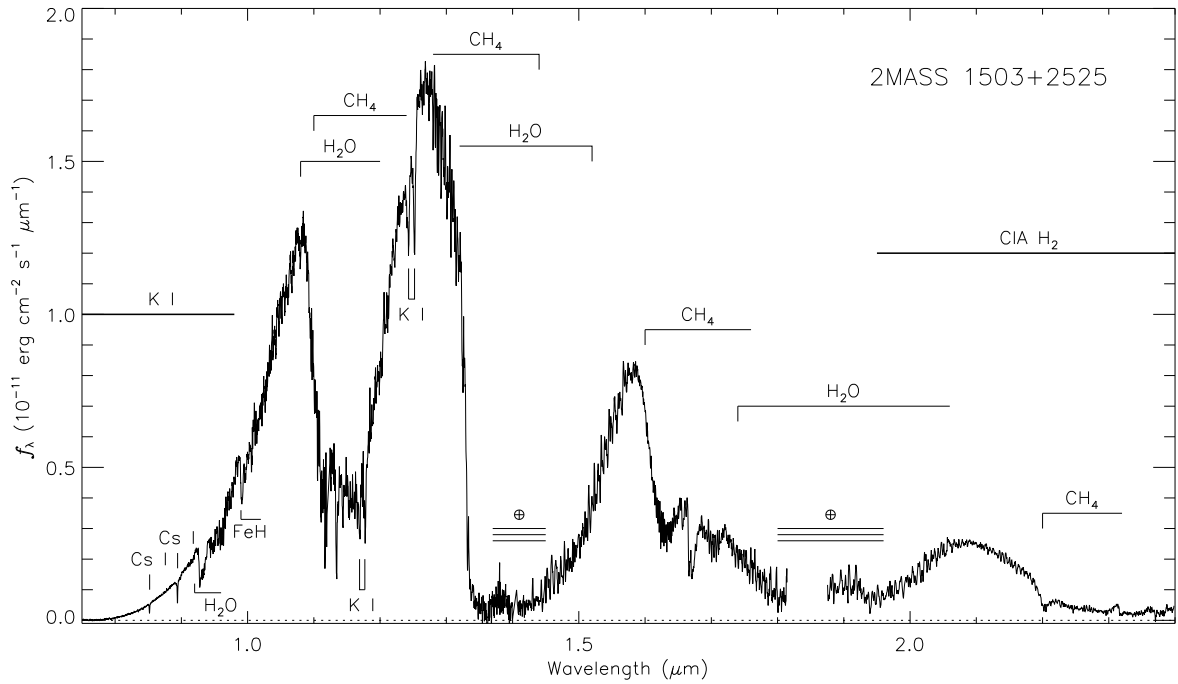}
\caption{The 0.75--2.5 $\micron$
spectrum of the T5 spectral
standard 2MASS 1503+2525 observed
at a spectral resolution {\ldl}$\sim$1200 \citep{me03d,me04a}.
Defining features of T dwarf near infrared spectra are labelled.
\label{fig_specfeatures}}
\end{figure}

Recently, two new spectral classes of low mass stars and brown dwarfs (stars with insufficient
mass to sustain hydrogen fusion; Kumar 1962; Hayashi \& Nakano 1963) have
been identified.  These sources, the
L dwarfs \citep{kir99,mrt99} and the
T dwarfs \citep[hereafter B02 and G02, respectively]{me02b,geb02},
lie beyond the standard stellar main sequence.
L dwarfs exhibit optical spectra with
waning TiO and VO bands (characteristic of M dwarfs); strengthening
metal hydride, alkali and H$_2$O absorption features; and increasingly
red optical/near infrared spectral energy distributions.
T dwarfs (Figure~\ref{fig_specfeatures})
are distinguished by the presence of CH$_4$ absorption
in their near infrared spectra, a species generally found in
planetary atmospheres \citep{geb96};
as well as pressure-broadened alkali resonance lines
in the optical, strong H$_2$O bands, and
collision induced H$_2$ absorption \citep{sau94,bor97} suppressing flux
at 2 $\micron$.
The sequence of spectral types from M to L to T is largely one
of decreasing effective temperature and luminosity
\citep[however, see $\S$~5.1]{dah02,vrb04,gol04},
and as such is a natural extension of the stellar main sequence.

Initial spectral classification schemes for the colder of these two classes,
the T dwarfs, have been proposed independently by B02 and G02.
Both are defined in the 1--2.5 $\micron$ spectral window
where T dwarfs emit the majority of their flux (e.g., Allard et al.\ 2001).
The underlying philosophies of these two schemes are somewhat different, however.
B02 identified seven representative standards spanning types T1 to T8
(excluding subtype T4), and classified a sizeable but inhomogeneous sample of low
and moderate resolution spectra by the comparison of a set of spectral indices.
G02 analyzed a smaller but homogenous
sample of {\ldl} $\sim 400$ near infrared spectra, and also
determined classifications using spectral indices, but
no standards were explicitly defined.

Neither of these schemes adhere rigorously to the MK Process;
yet, despite their underlying differences, classifications differ
by no more than 0.5 subtypes \citep{me03x}.
On the other hand, the existence of two separate
classification schemes for T dwarfs has led to
redundancies and confusion in the literature
(e.g., Scholz et al.\ 2003).  Furthermore, the relatively few T dwarfs
(roughly 30) known at the time these schemes were introduced,
and the limited observations available for them,
resulted in an incomplete sampling of the T class
and a potentially biased set of spectral standards
(e.g, contaminated by peculiar sources and unresolved multiple systems).
With over twice as many T dwarfs now known, primarily identified in
the Sloan Digital Sky Survey \citep[hereafter SDSS]{yor00}
and the Two Micron All Sky Survey \citep[hereafter 2MASS]{cut03},
and with extensive imaging, spectroscopic and astrometric data now available,
it is an opportune time to revisit the classification of these cold brown dwarfs.

In this article, we present a revised near infrared spectral classification
scheme for T dwarfs which unifies and supersedes the
studies of B02 and G02.  This scheme is defined by a set of carefully screened
spectral standard stars spanning types T0 to T8,
and is demonstrated on two distinct, large and homogenous spectral samples.
In $\S$~2 we describe the primary spectral libraries employed for this study,
as well as additional published datasets examined.
In $\S$~3 we identify and give detailed information
on the nine primary and five alternate spectral
standards used to define the sequence.
In $\S$~4 we review the methods of classifying T dwarfs,
focusing first on
the direct comparison of spectral data to the standard spectra
as dictated by the MK Process, then describing
secondary methods based on redefined spectral indices
sampling the major near infrared {\water} and {\meth} bands.
In $\S$~5 we discuss the revised subtypes, comparing them
to prior classifications.  We also
discuss how extensions to this one-dimensional
scheme may be made to account for secondary spectral variations,
arising from physical differences in atmospheric dust content,
surface gravity and metallicity; and speculate on the end of the T dwarf class.
Individual sources are discussion in $\S$~6.
Results are summarized in $\S$~7.
We provide a compendium of all presently known T dwarfs
and their revised spectral types in the Appendix.

\section{Spectral Data}

Defining the classification of a stellar class necessitates
a sizeable set of homogeneous (similar resolution and wavelength coverage)
spectra for both standards and classified sources.
Our study draws primarily on two large, near infrared spectral
libraries of T dwarfs obtained with the SpeX instrument \citep{ray03},
mounted on the 3.0m NASA Infrared Telescope
Facility Telescope; and the Cooled Grating Spectrometer 4 \citep[hereafter CGS4]{wri93},
mounted on the 3.5m United Kingdom Infrared Telescope (UKIRT).
These libraries are available in electronic form upon request.\footnote{Also see
\url{http://DwarfArchives.org} and \url{http://www.jach.hawaii.edu/{$\sim$}skl/LTdata.html}.}

The SpeX dataset is composed of prism-dispersed spectra
covering 0.8-2.5 $\micron$ in a single order
at a spectral resolution {\ldl} $\approx$ 150.
These low resolution data sufficiently sample the broad H$_2$O and CH$_4$ bands
present at near infrared wavelengths, but cannot resolve important
line features such as the 1.25 $\micron$
\ion{K}{1} doublets (Figure~\ref{fig_specfeatures}).
The SpeX sample includes 59 spectra of 43 T dwarfs and two
optically-classified L8
dwarfs \citep{kir99}, a subset
of which have been previously published \citep{me04a,me05a,cru04}.
All data have been homogeneously acquired and reduced using the
Spextool package \citep{vac03,cus04}.

The CGS4 dataset is composed of 41
spectra of 39 T dwarfs and two optically-classified L8 dwarfs, with
typical resolutions {\ldl} $\approx$ 300-500.\footnote{With the
exception of the bright T dwarf
2MASS 0559$-$1404, which was observed at twice this resolution
(G02).} At these slightly higher resolutions, details
within the molecular bands and atomic line absorptions can be resolved.
The CGS4 spectra, nominally spanning 0.85--2.5 $\micron$, require four instrumental
settings to acquire,
although some of the data encompass a subset of this spectral range.
Nearly all of these spectra have been previously
published \citep{geb96,str99,leg00,tsv00,geb01,geb02,kna04}, and data acquisition and
reduction procedures can be found in the literature.

In addition to these two primary datasets,
we have examined other late-type L and T dwarf spectra
reported in the literature
\citep{opp95,cub99,me00c,me02a,me03e,nak01,nak04,liu02,zap02,mcl03,mcc04,cus05}.
As these data have been obtained
with assorted instrumentation, they vary in both
spectral resolution and wavelength coverage.
Details of all of the spectral
datasets examined here are given in Table~\ref{tab_speclib}.

While many of the T dwarfs with near infrared
spectral data have been observed by multiple instruments
(e.g., 25 have been observed with both SpeX and CGS4), nearly all
have had no more than two separate observations with a single instrument.
The general absence of spectral monitoring observations
prevents a robust analysis of spectral variability
for T dwarfs and its impact on their classification.
We therefore assume that the spectra are
effectively static and representative of each source over long periods.

\section{Spectral Standards}

\subsection{Primary Standards}

The selection of spectral standards is the most important
aspect of defining a classification scheme, as these sources
provide the framework for the entire class.
The set of standards should encompass the full range of spectral morphologies
observed while being sufficiently unique so as to
be readily distinguishable.  Peculiar (e.g., unusual metallicity) or
highly variable standards
are poor choices as they may improperly skew the sequence.
A standard that is too difficult to observe --- due to its faintness, unobservable
declination or obscuration by a nearby bright star --- is also of limited utility.

We have therefore attempted to select T dwarf spectral standards
that conform to the following criteria:
\begin{itemize}
\item reasonably bright,
\item not known to be spectroscopically peculiar (see $\S$~6),
\item not known to be a resolved multiple system and
\item within 25$\degr$ of the celestial equator.
\end{itemize}
We have considered both previously selected standards (B02) and
more recently discovered sources for which extensive data (spectroscopic
and otherwise) have been obtained.
In this manner, nine primary standards
spanning subtypes T0 through T8 that best
represent the known population of T dwarfs were identified.
Coordinates and photometric measurements
are given in Table~\ref{tab_standards}; additional
data are provided in the Appendix.  Detailed descriptions are as follows:

\noindent {\it SDSS 1207+0244\footnote{Source designations are
abbreviated in the manner SDSS~hhmm$\pm$ddmm, where
the suffix conforms to IAU nomenclature convention and is the sexigesimal Right
Ascension (hours and minutes) and declination (degrees and arcminutes)
at J2000 equinox. Full designations are provided for all known T dwarfs
in the Appendix.} (T0):} Recently identified in the SDSS by \citet{kna04}
and classified T0 on the G02 scheme,
this source in favored over the bright, unequal-brightness binary SDSS 0423$-$0414AB
(G02; Burgasser et al.\ 2005).
No high resolution imaging or parallax observations have
yet been made for this source.

\noindent {\it SDSS 0837$-$0000 (T1):}  One of the first
``L/T transition objects'' discovered by \citet{leg00},
this source was the T1 spectral standard
on the B02 scheme and classified T0.5 by G02.
It is unresolved in {\em Hubble Space Telescope} {\em (HST)}
observations (Burgasser et al.\ in prep.)
and has a poorly constrained parallactic distance of 29$\pm$12 pc \citep{vrb04}.

\noindent {\it SDSS 1254$-$0122 (T2):}  Also identified by \citet{leg00},
this relatively bright source ($J = 14.66{\pm}0.03$;
Leggett et al.\ 2000\footnote{Near infrared photometry
reported in the text are generally based on the Mauna Kea Observatory (MKO) filter system
\citep{sim02,tok02}, unless otherwise specified (e.g., Table~\ref{tab_allt}).}) was the
T2 standard in the B02 scheme and classified likewise on the G02 scheme.
It is a single source in {\em HST} images (Burgasser et al.\ in prep.).  Three parallax
distance measurements have been made for SDSS 1254$-$0122, one in the
optical (11.8$\pm$0.3 pc, Dahn et al.\ 2002) and two in the near infrared
(13.7$\pm$0.4, Tinney et al.\ 2003; and 13.2$\pm$0.5 pc, Vrba et al.\ 2004);
note the disagreement.  A more precise measure is expected from the USNO
near infrared parallax program (F.\ Vrba 2005, priv.\ comm.).

\noindent {\it 2MASS 1209$-$1004 (T3):}  This recently discovered T dwarf \citep{me04a}
replaces the apparent double SDSS 1021$-$0304AB (Burgasser et al.\ in prep.)
as the T3 standard on the B02 system.  No high resolution imaging or parallax
measurements of this source have yet been obtained.

\noindent {\it 2MASS 2254+3123 (T4):}  While outside of our declination
constraint, the spectrum of this relatively bright ($J$ = 15.01$\pm$0.03;
Knapp et al.\ 2004) T dwarf fits ideally between those of
our T3 and T5 standards.  Identified by B02 and originally
classified T5 on that scheme (Knapp et al.\ 2004 classify it T4
on the G02 scheme), it is unresolved in {\em HST} observations (Burgasser et al.\ in prep.).
No parallax measurement has been reported for this source.
\citet{eno03} report a marginally significant rise of 0.5$\pm$0.2 mag
in the $K$-band flux of this object over
the course of three nights, but this possible detection of
variability has yet to be confirmed.

\noindent {\it 2MASS 1503+2525 (T5):}  This bright source
($J = 13.55{\pm}0.03$; Knapp et al.\ 2004)
was identified by \citet{me03a} and originally classified T5.5 on the B02
scheme. While at a slightly higher
declination than our adopted selection criteria, the brightness of 2MASS 1503+2525
and lack of a visible companion (Burgasser et al.\ in prep.)
make it an excellent choice as a spectral standard. No parallax
distance measurement has been reported for this source.

\noindent {\it SDSS 1624+0029 (T6):}  The first known field T dwarf, identified by \citet{str99}
in the SDSS database and classified T6 on both the B02 and G02 schemes,
SDSS 1624+0029 is a representative and easily
accessible standard.  It is unresolved in {\em HST}
imaging observations (Burgasser et al.\ in prep.), and
has a parallax distance measurement of 11.00$\pm$0.15 pc
\citep[see also Dahn et al.\ 2002; Vrba et al.\ 2004]{tin03}.  \citet{nak00}
report very low levels (1-3\%) of variability in fine {\water}
features between 1.53 and 1.58 $\micron$ in the spectrum of the source,
but not significant enough to affect
its gross spectral morphology.

\noindent {\it 2MASS 0727+1710 (T7):}  Identified by B02 and selected as the T7
standard in that scheme (Knapp et al.\ 2004 classify it T8 on the G02 scheme),
this source remains an excellent spectral standard.  No high resolution imaging
observations have been reported for 2MASS~0727+1710, but it has
a parallax distance measurement of 9.09$\pm$0.17 pc \citep{vrb04}.

\noindent {\it 2MASS 0415$-$0935 (T8):}  The coldest ({\teff} $\approx$ 700 K;
Vrba et al.\ 2004; Golimowski et al.\ 2004) and latest-type T dwarf known,
this source was initially identified by B02 and selected as the T8 standard
in that scheme.  It is the sole T9 on the G02 system \citep{kna04}.
2MASS 0415$-$0935 is unresolved in {\em HST} imaging observations (Burgasser et al.\ in prep.), and
is the closest (isolated)
field T dwarf to the Sun currently known,
with a parallax distance measurement of 5.75$\pm$0.10 pc
\citep{vrb04}.

\begin{figure}
\epsscale{1.0}
\plottwo{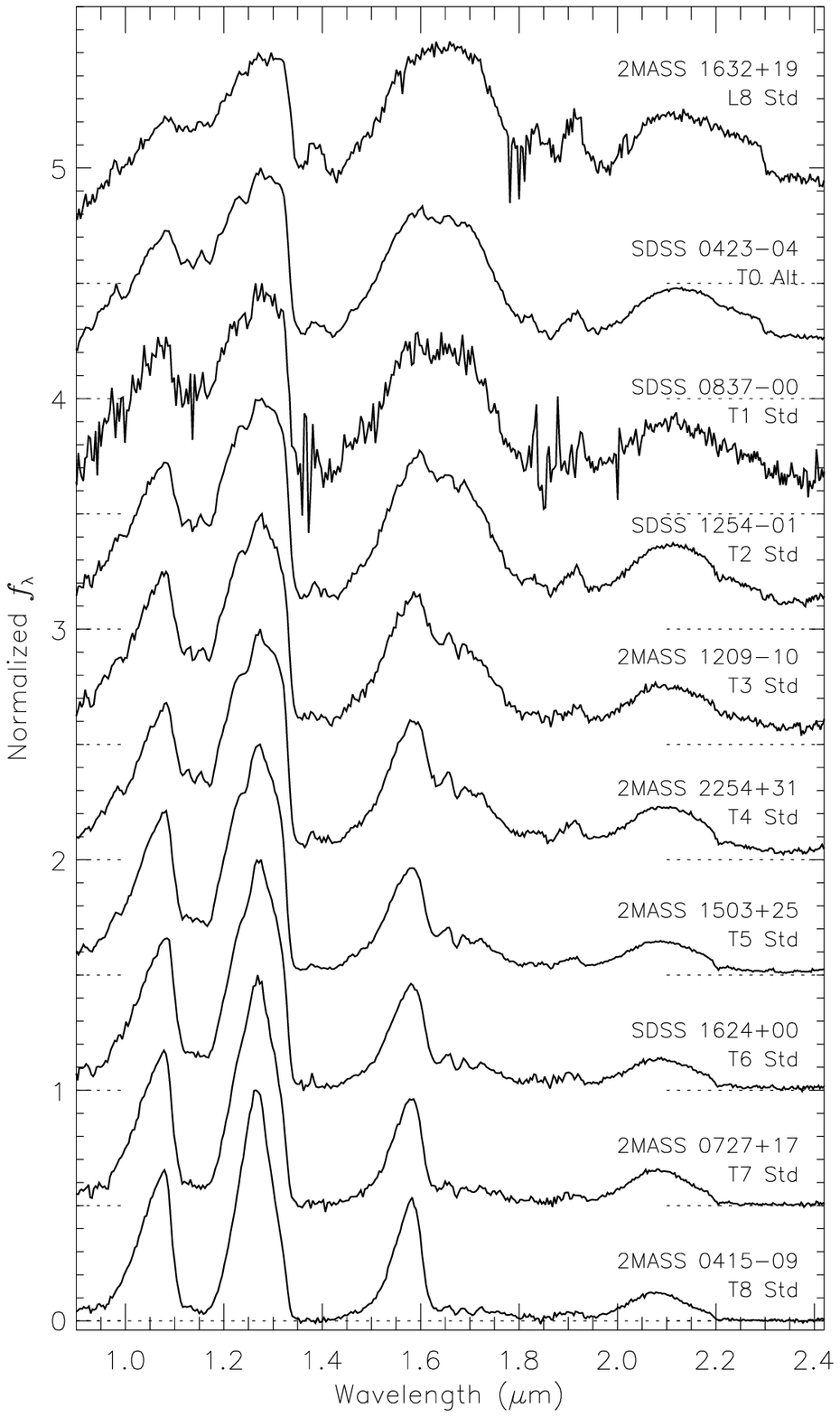}{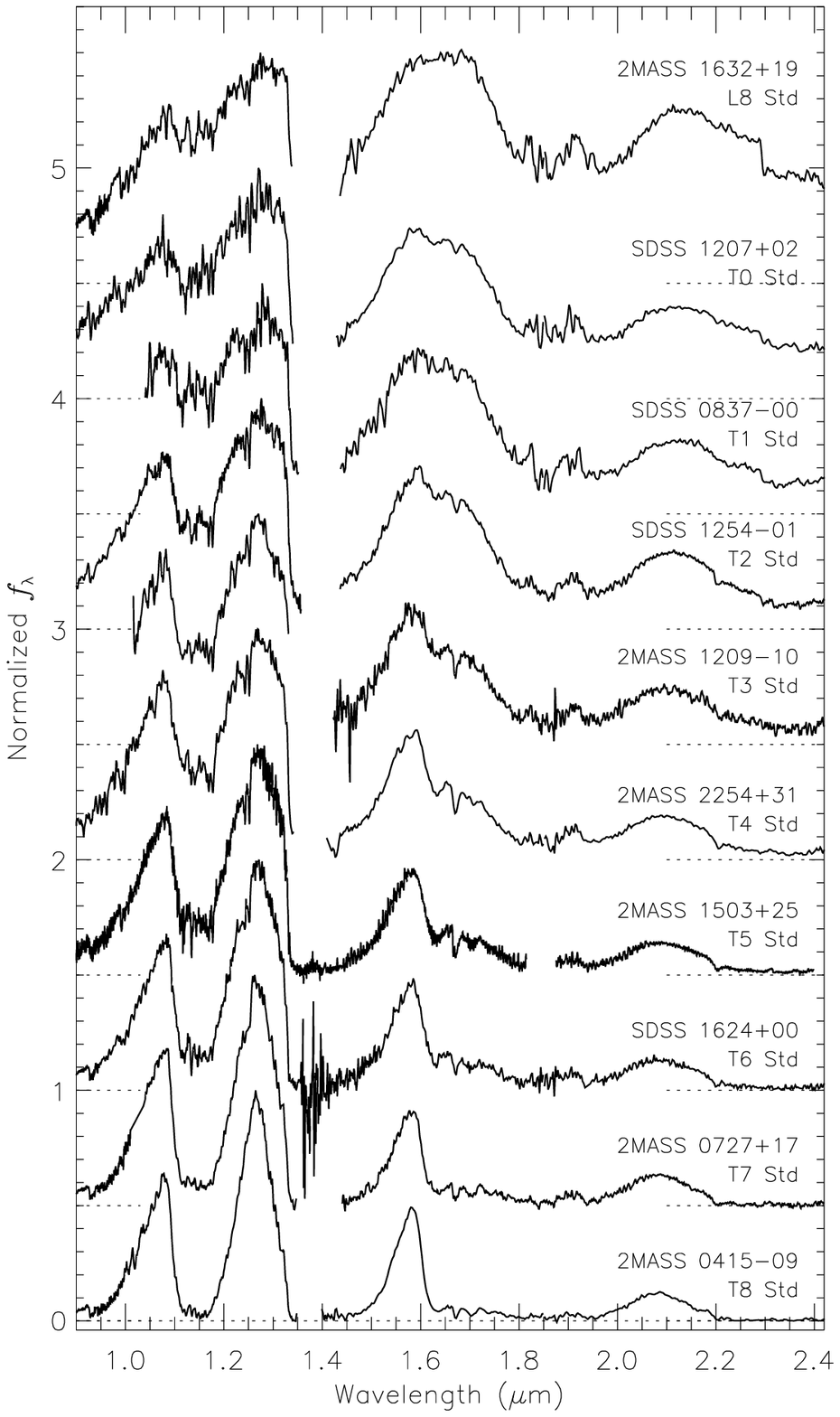}
\caption{Near infrared spectra of T dwarf standards, along with the
L8 optical standard 2MASS 1632+1904 \citep{kir99}.  Left panel displays the
low resolution SpeX sample (note the substitution of the alternate T0 standard
SDSS 0423$-$0414AB), right panel displays the moderate
resolution CGS4 sample (with SpeX cross dispersed data for 2MASS 1503+2525).
All spectra are normalized at 1.25 $\micron$
and offset by a constant (dotted lines).
\label{fig_standards}}
\end{figure}

Figure~\ref{fig_standards} displays the spectral sequence of these standards, along with
the L8 optical standard 2MASS 1632+1904 \citep{kir99}, for both the SpeX
and CGS4 datasets.  These spectra effectively define the T dwarf class.
The emergence of $H$-band {\meth} absorption at these spectral
resolutions defines the start of the T dwarf sequence,
as originally proposed by G02.
Early-type T dwarfs exhibit weak {\meth} bands, strong {\water} bands
and waning CO absorption at
2.3 $\micron$. In later types,
{\water} and {\meth} bands progressively strengthen;
the 1.05, 1.25, 1.6 and 2.1 $\micron$ peaks become more pronounced and acute;
and the $K$-band peak becomes increasingly suppressed relative to $J$.
The end of the T class is exemplified by the spectrum of 2MASS 0415$-$0935,
with nearly saturated {\water} and {\meth} bands, and sharp triangular
flux peaks emerging between these bands.  The range of spectral morphologies
encompassed by the standards in Figure~\ref{fig_standards} span the full
spectral variety of the currently known T dwarf population.

\subsection{Alternate Standards}

In addition to the primary standards, we have identified a handful of alternate standards
that have nearly identical near infrared spectral energy distributions
but are well-separated on the sky.  While in some cases
these sources do not strictly adhere to the constraints outlined above,
their purpose is to facilitate the
observation of a spectral comparator at any time of the year.
The alternate standards are
listed in Table~\ref{tab_standards} and described as follows:

\noindent {\it SDSS 0423$-$0414AB (T0 alternate):}
This relatively bright source ($J = 14.30{\pm}0.03$; Leggett et al.\ 2002b)
was identified by G02. Its spectrum, like that of SDSS 1207+0244,
exhibits exceedingly weak CH$_4$ absorption
at $H$-band and both CO and CH$_4$ bands at
$K$-band.  SDSS 0423$-$0414AB was not selected as a primary standard here, however,
as {\em HST} observations resolve it as an unequal-brightness binary \citep{me05c}.
Nevertheless, its near infrared spectrum matches that of the T0 standard.
SDSS 0423$-$0414AB has a parallax distance measurement of 15.2$\pm$0.5 pc \citep{vrb04}.

\noindent {\it SDSS 0151+1244 (T1 alternate):} Identified by G02, this faint source
($J$ = 16.25$\pm$0.05; Leggett et al.\ 2002b) is classified T1 on the G02 scheme.
It is unresolved in {\em HST} images (Burgasser et al.\ in prep), and has
a parallax distance measurement of 21.4$\pm$1.6 pc \citep{vrb04}.
\citet{eno03} report significant $K$-band variability from this source
(0.42$\pm$0.14 mag peak-to-peak) with a possible period of 2.97 hours.
The faintness and apparent variability of SDSS 0151+1244 relegate it as
an alternate standard, although its
near infrared spectrum is very similar to that of SDSS 0837$-$0000.

\noindent {\it SDSS 1021$-$0304AB (T3 alternate):} Identified by \citet{leg00}
and resolved as an unequal-brightness binary in {\em HST} observations
(Burgasser et al.\ in prep), this original T3 standard on the B02
scheme (also classified T3 by G02)
remains a viable alternative to the primary standard 2MASS 1209$-$1004,
although it is only separated by 28$\degr$ on the sky.
SDSS 1021$-$0304AB has a parallax distance measurement of 29$\pm$4 pc
\citep[see also Vrba et al.\ 2004]{tin03}.

\noindent {\it 2MASS 0755+2212 (T5 alternate):} Identified by B02 and classified T5 on
that system, this source has a nearly identical spectrum to that of
the brighter T5 standard 2MASS 1503+2525 (Figure~\ref{fig_spexviscomp}).
It is unresolved in {\em HST} images (Burgasser et al.\ in prep).
No parallax measurement has been reported.

\noindent {\it 2MASS 1553+1532AB (T7 alternate):} This late-type T dwarf identified
by B02 and classified T7 on that scheme
has a near infrared spectrum identical to that of the T7 primary standard
2MASS 0727+1710.
Resolved into an equal-magnitude, co-moving pair with
{\em HST} and ground-based observations (Burgasser et al.\ in prep),
2MASS 1553+1532AB appears to be composed of two T dwarfs with similar
spectral types.  No parallax measurement has been reported for this source.

\section{Methods of Classification}

\subsection{Direct Spectral Comparison}

The most straightforward and accurate means of classifying any stellar spectrum
is through the direct comparison of that spectrum to an
equivalent set of spectral standards.  Direct spectral comparison
enables the simultaneous examination of multiple features,
providing a broad match to the overall spectral morphology.
Furthermore, this method facilitates the identification of peculiar sources
which do not fit within the standard sequence,
possibly due to secondary parameters or the presence of an unresolved
companion (see $\S$~6).  Direct spectral comparison is the proscribed means
of classification via the MK Process \citep{mor73}.

\begin{figure}
\epsscale{1.0}
\plotone{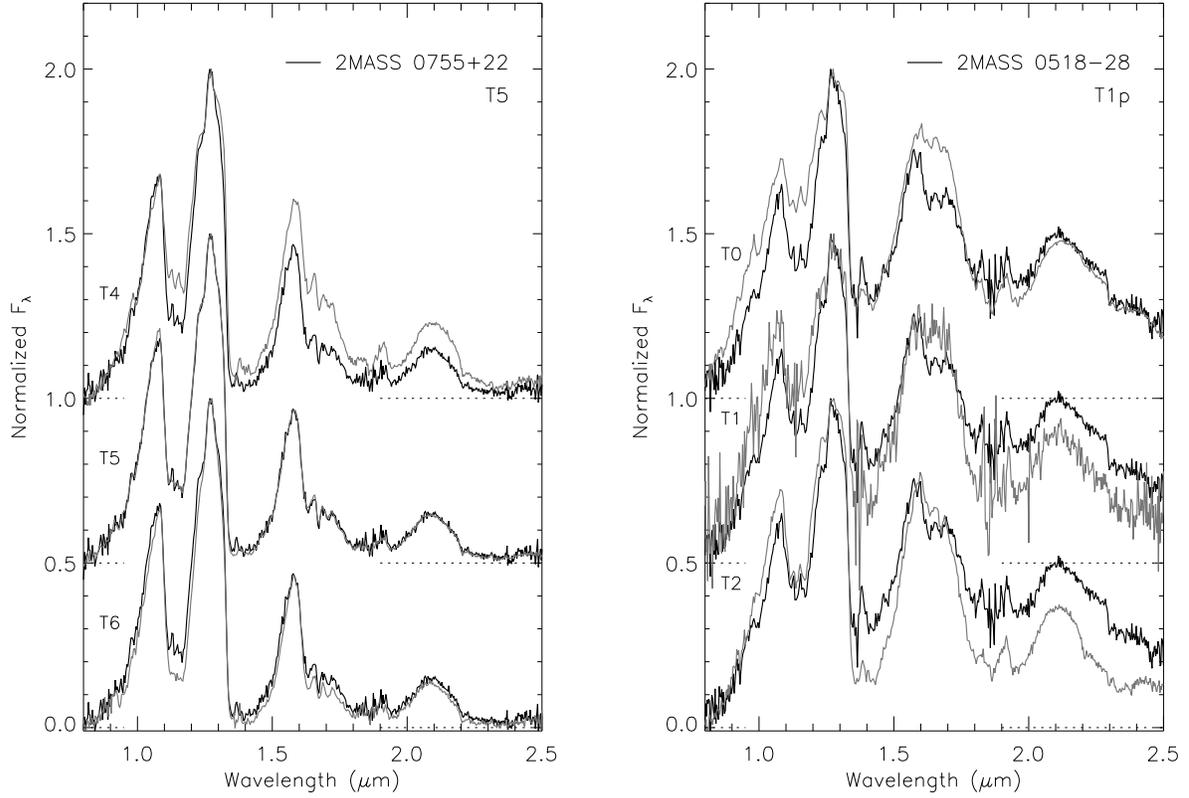}
\caption{Examples of direct spectral classification
for the low resolution SpeX data.  The left panel demonstrates
the case of the T5 2MASS 0755+2212 (B02), which closely matches
the T5 standard 2MASS 1503+2525.  The right panel demonstrates the
case of the peculiar T1 2MASS 0518-2828 \citep{cru04}, the spectrum of which does
not match any of the T dwarf standards.  This source is suspected
to be an unresolved L dwarf/T dwarf binary.  In both panels, source (black) and
standard (grey) spectra
are normalized at their $J$-band flux peaks and offset by constants
(dotted lines).
\label{fig_spexviscomp}}
\end{figure}

\begin{figure}
\epsscale{1.0}
\plotone{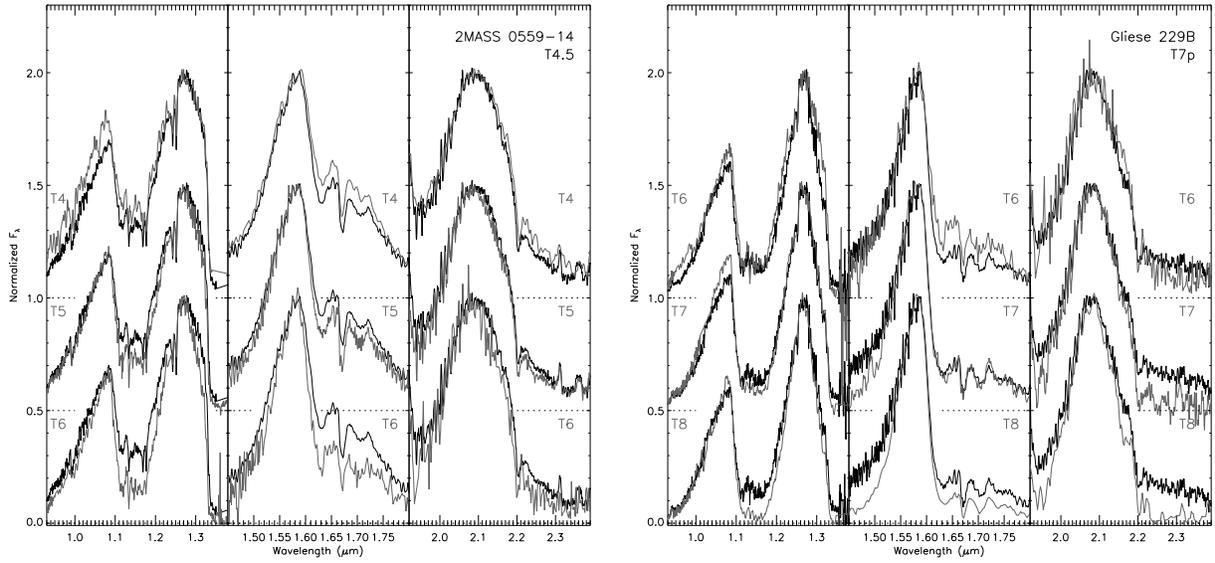}
\caption{Similar to Figure~\ref{fig_spexviscomp} for moderate resolution CGS4 data.
Here we show the examples of the T4.5 2MASS 0559$-$1404 and the peculiar T7
Gliese 229B.  Spectral data are separated into
three bands, with the source (black) and standard (grey) spectra normalized at
their flux peaks within the band.  Note how the features in the
spectrum of 2MASS 0559$-$1404 fall midway in strength between the T4 and T5 standard; while
the spectrum of Gliese 229B is inconsistent with any of the standards
in all three spectral regions simultaneously.
\label{fig_cgs4viscomp}}
\end{figure}

Spectral types for all of the T dwarfs in the SpeX and CGS4 datasets
were determined by overlaying their normalized spectra
onto the corresponding T dwarf standard sequences\footnote{With two exceptions: we
used the alternate standard SDSS 0423$-$0414AB as the T0 comparator
for the SpeX dataset, and SpeX cross-dispersed data were used for
the T5 standard 2MASS 1503+2525 in the CGS4 dataset.}
shown in Figure~\ref{fig_standards}.
The standard sequences were augmented with
near infrared data for the L8
optical standard 2MASS 1632+1904 \citep{kir99}; and
for 2MASS 0310+1628 \citep{kir00},
classified L9 in the near infrared by G02.
Both source and standard spectra were normalized
and compared in a consistent manner depending on the dataset.
For the low resolution SpeX data,
the entire 0.8--2.5 $\micron$ range was examined simultaneously
after normalizing at the $J$-band flux peak.
For the CGS4 data, spectra were separately normalized and
compared in three wavebands ---
0.95--1.35, 1.45--1.8 and 1.9--2.4 $\micron$ --- to minimize
errors in order scaling and to discern finer details.
Examples of each of these methods is shown in Figures~\ref{fig_spexviscomp}
and~\ref{fig_cgs4viscomp}.

In both cases, subtypes for individual T dwarfs were assigned according
to which standard spectra provided the closest spectral match.  This was gauged by
the relative strengths of the major {\water} and {\meth} bands, detailed
shapes of the flux peaks and overall spectral slope.
Half-subtypes were assigned for those spectra that
clearly fell between standards (e.g., 2MASS 0559$-$1404 in Figure~\ref{fig_cgs4viscomp}).
While this comparative technique is qualitative in nature, the distinct
spectral morphologies of the standard spectra enabled unambiguous
classifications in nearly all cases.
As the sequence is set up in whole subtype intervals, the nominal
precision of the scheme is 0.5 subtypes for reasonable signal-to-noise (S/N) spectra.
Classifications based on low S/N spectra are more uncertain and were
specifically noted as such.

In some cases, the spectrum of a source does not fall anywhere in the
standard sequence due to conflicting band strengths or substantially
inconsistent broad band colors (the latter discernable with the SpeX prism data).
These peculiar sources are exemplified in Figures~\ref{fig_spexviscomp}
and~\ref{fig_cgs4viscomp}
for the cases of 2MASS 0518$-$2828 \citep{cru04} and Gliese 229B \citep{nak95,geb96},
respectively.  Peculiar T dwarfs were specifically
labelled and are discussed in further detail in $\S$~6.

Tables~\ref{tab_spexclass} and~\ref{tab_cgs4class} (column 9)
list the resulting subtypes as determined by direct
comparison for the SpeX and CGS4 sources, respectively.
Subtypes for other data reported in the literature
were also determined by comparing those
spectra to standards obtained at similar resolution; e.g., Keck NIRC
data were compared to the SpeX standards, while Keck NIRSPEC
(low resolution) data were compared to the CGS4 standards.  Derived
spectral types for these data
are given in Tables~\ref{tab_nircclass}--\ref{tab_litclass}.
In all cases, the classifications of sources with multiple
spectral data
are identical within the scheme's 0.5 subclass precision.
The consistency of the classifications demonstrates the reliability of the
direct spectral comparison technique.

\subsection{Classification by Spectral Indices}

In cases where one is dealing with
large spectral samples or desires a quantification of deviations from the standard
system (e.g., when dealing with dust or gravity effects; see $\S$~5.2),
measurements of diagnostic features through spectral
indices can be useful.  Such indices, typically defined as the ratio of
spectral flux or flux density in two different wavebands, have had widespread use in the
classification of late-type dwarfs, whose spectra typically have
strong molecular and atomic features.  For example, optical TiO and CaH indices
defined by \citet{rei95} and employed by \citet{giz97} are used to
segregate solar metallicity and metal-poor
M dwarfs; while near infrared indices have been widely employed in the classification
of M and L dwarfs \citep{jon94,jon96,del97,del99,tin98,tok99,rei01,tes01}.
Both B02 and G02 make use of spectral indices in their classifications, sampling
the major H$_2$O and CH$_4$ bands, the 0.8--1.0 $\micron$ red slope, color ratios and
the detailed shapes of the $J$- and $K$-band flux peaks.  We revisit a subset of these indices here.

\subsubsection{Revised Spectral Indices}

The 1.1, 1.4 and 1.8 $\micron$ H$_2$O and 1.3, 1.6 and 2.2 $\micron$ CH$_4$
absorption bands are the most
dominant and defining features of T dwarf spectra.  These bands
vary significantly and monotonically
throughout the standard sequence and are generally correlated --- strong H$_2$O bands
are generally associated with strong CH$_4$ bands.
These bands are also broad enough
to be measured at low spectral resolutions, and are found
in multiple spectral regions throughout the 1.0--2.5 $\micron$ range.
Because of their utility, our spectral index analysis
is focused on these strong molecular features.

We reexamined the H$_2$O and CH$_4$ indices defined in B02 and G02 in order
to optimize their use with T dwarf spectra.
They are redefined here as the ratio of the integrated flux
over a spectral window within an absorption feature to the
integrated flux over the same-sized window in the neighboring
pseudocontinuum.  This definition minimizes large
fluctuations arising from poor S/N at the bottom of
strong bands (e.g., in the latest-type T dwarfs).
We also attempted to avoid regions of strong telluric absorption
in order to minimize variations between spectra obtained at
different sites.
Finally, the spectral ratio wavebands were defined to be
broad enough to facilitate use with both low and moderate spectral resolution data.

\begin{figure}
\epsscale{1.0}
\plotone{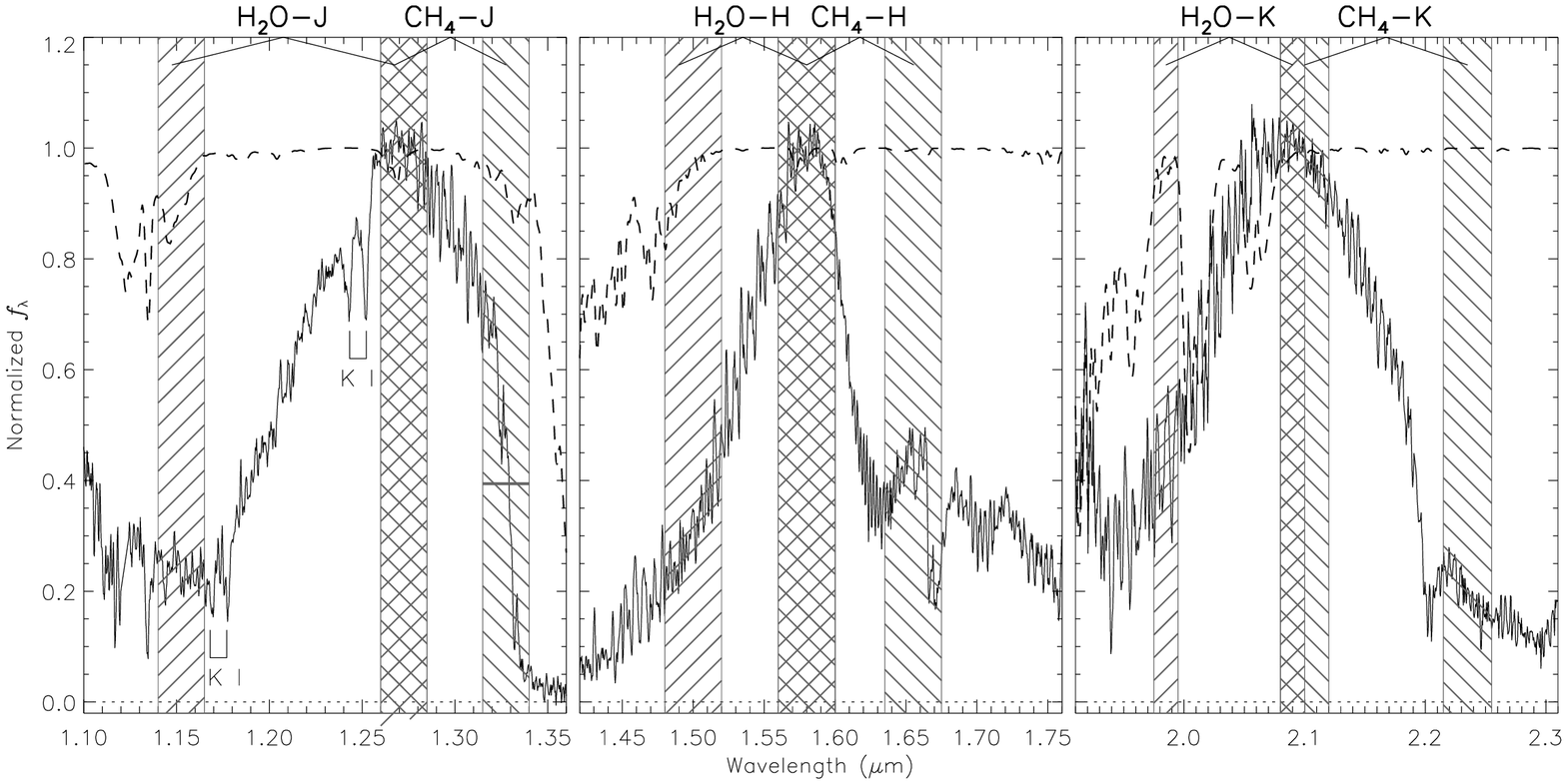}
\caption{Spectral regions sampled by the six {\water}
and {\meth} indices
defined in Table~\ref{tab_indices}, plotted on the
near infrared spectrum of the T5 standard 2MASS 1503+2525 \citep{me04a}.
Not shown are the regions sampled by the K/J index.
A telluric transmission spectrum typical for Mauna Kea is also shown
(dashed lines) to highlight regions of low terrestrial atmospheric absorption.
Spectral data in each panel are normalized at the local flux peak.
\label{fig_indspec}}
\end{figure}

Table~\ref{tab_indices} lists the six revised
classification indices, along with a seventh index measuring the ratio of flux
between the $K$- and $J$-band flux peaks.  This last index samples the relative
strength of H$_2$ absorption, a gravity and metallicity-sensitive feature (e.g., B02),
and is discussed in further detail in $\S$~5.2.
Figure~\ref{fig_indspec} diagrams the passbands of the {\water} and {\meth} indices
on the spectrum of the T5 standard 2MASS 1503+2525 and a telluric absorption
spectrum typical of Mauna Kea.\footnote{These data, produced using the program IRTRANS4,
were obtained from the UKIRT worldwide web pages.}

\begin{figure}
\epsscale{0.8}
\plotone{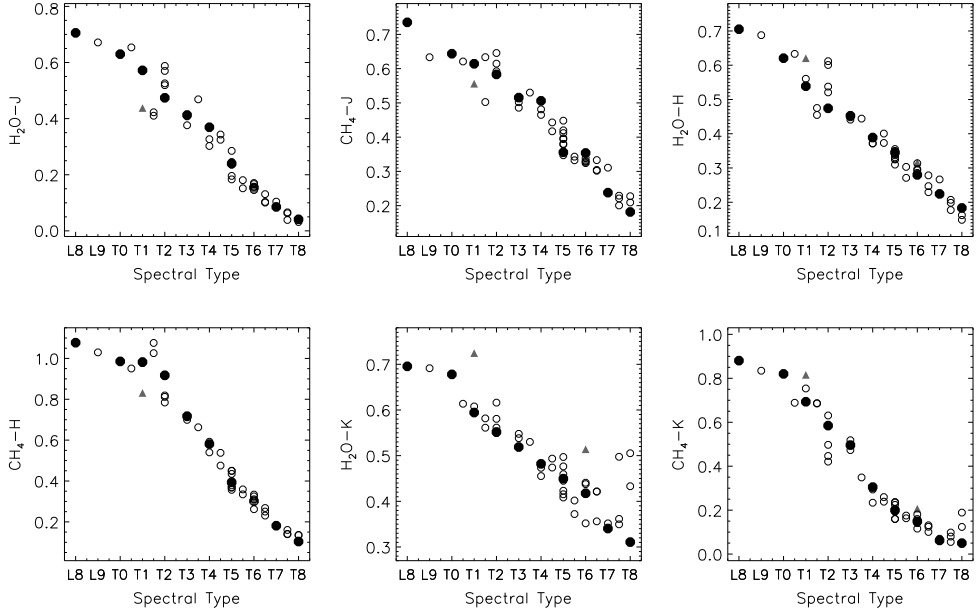}
\caption{Values for the {\water} and {\meth} spectral indices measured on
the SpeX prism data as a function of
spectral type (as determined by direct spectral comparison).
Primary standards are indicated by solid black circles, peculiar
sources and uncertain classifications by solid grey triangles, and all other sources by
open circles.  Note the variable vertical scale in each panel.
\label{fig_spexind}}
\end{figure}

\begin{figure}
\epsscale{0.8}
\plotone{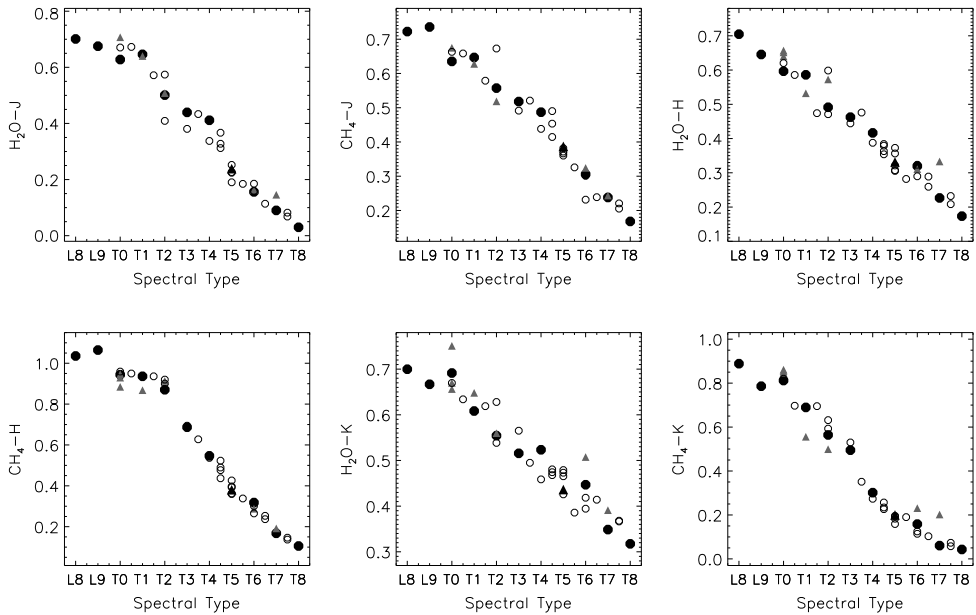}
\caption{Same as Figure~\ref{fig_spexind} for the CGS4 spectra.
\label{fig_cgs4ind}}
\end{figure}

Measurements of the redefined indices for spectral data examined here
are given in Tables~\ref{tab_spexclass}--\ref{tab_litclass} (columns 2--8).
In Figures~\ref{fig_spexind} and~\ref{fig_cgs4ind}, we plot the
values of the {\water} and {\meth} indices as measured with the SpeX and CGS4 data,
respectively, as a function of spectral type.
All six ratios show monotonic decreases in value with later subtypes,
with strong correlations.  In particular, the
{\water}-H index varies nearly linearly over the entire
spectral type range.  The {\water}-J, {\meth}-J and
and {\meth}-H indices
are slightly degenerate for the earliest T spectral
types, while the {\meth}-K index is degenerate between T7 and T8.
The index {\water}-K shows the largest scatter as a function of spectral
type, notably among late-type and peculiar sources.  Indeed, for some datasets
(e.g., OSIRIS), there is very poor correlation between this index and spectral type,
likely due to increased noise caused by telluric absorption around the 1.9 $\micron$
{\water} band.  We therefore omit this index from our classification set,
and focus on five primary classification indices: {\water}-J, {\meth}-J,
{\water}-H, {\meth}-H and {\meth}-K.

\subsubsection{Index Classification: Two Methods}

The B02 and G02 schemes differ slightly in their use of spectral indices for
classification.  The former scheme directly compares
measured indices to those of the standards, while the latter scheme
specifies subtype ranges for each index.
The B02 method is more closely aligned to the MK Process in spirit, but the
G02 method is a less cumbersome means of determining the classification of a
particular T dwarf.
As both of these techniques have been used in the literature,
we examined them independently to determine whether any significant
differences exist.

Index subtypes were first derived following the prescription of B02.
Table~\ref{tab_indstandard} lists the spectral
index values for the T dwarf standards
with SpeX and CGS4 data, in addition to values for the L8 and L9 comparison
stars 2MASS 1632+1904 and
2MASS 0310+1648.  As in B02, individual
index subtypes were first determined as the closest match
to the standard values (or half-subtype match).  Subtypes for the
{\meth}-H index were only used for index values less than 1.0,
as higher values are degenerate for L dwarfs and the earliest T
dwarfs.  Final classifications were
determined as the average of the index subtypes; no rejection of outliers was made
as in B02.
Sources with a large scatter amongst the index subtypes ($>$ 1 subclass)
or with only one spectral index measure available were noted as uncertain.

The second method, following the prescription of G02, involves the comparison of
spectral indices to predefined ranges.  Table~\ref{tab_indrng} lists these ranges
for the new indices,
determined as the typical values for each spectral type as measured on
the CGS4 data (i.e., Figure~\ref{fig_cgs4ind}).  Because of their degeneracy,
ranges for the {\water}-J and {\meth}-K indices are only defined for spectral
types $\geq$T2 and $\leq$T6, respectively.  Index subtypes for each T dwarf
were assigned according to the range the value falls in (or half subtype for values close
to range borders), and the average type was determined in the manner described above.

\begin{figure}
\epsscale{0.8}
\plotone{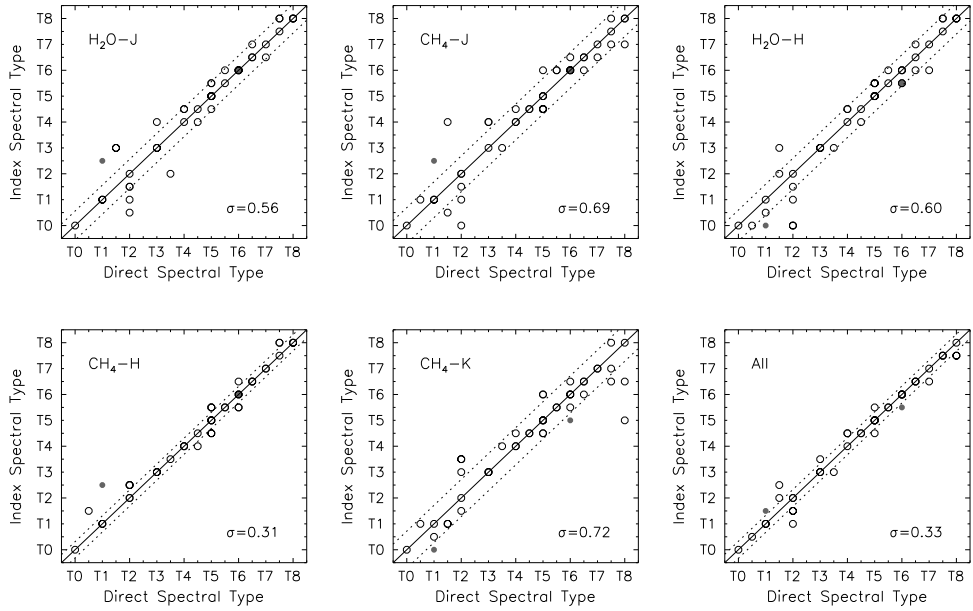}
\caption{A comparison of index subtypes for the five primary
classification indices, computed using the B02 method,
versus subtypes determined by direct spectral comparison
for the SpeX prism data.  Peculiar T dwarfs and uncertain classifications
are indicated by
solid grey circles, all others by open circles.  The solid line
delineates perfect agreement between subtypes
while the dotted lines indicate the $\pm$1$\sigma$
scatter in subtype deviations, as listed in each panel.
\label{fig_stdindvsspt_spex}}
\end{figure}

\begin{figure}
\epsscale{0.8}
\plotone{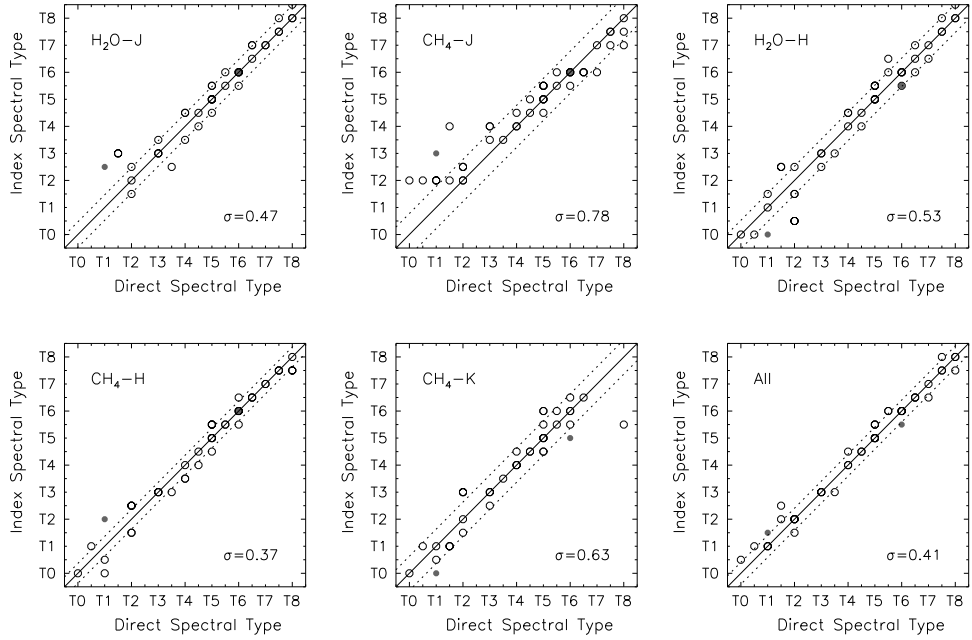}
\caption{Same as Figure~\ref{fig_stdindvsspt_spex}, but comparing index subtypes
computed using the G02 method.  Note the absence of data points where the
index ranges (Table~\ref{tab_indices}) are not defined.
\label{fig_rngindvsspt_spex}}
\end{figure}

Tables~\ref{tab_spexclass}--\ref{tab_litclass}
list the index-based spectral subtypes for the five primary classification
ratios (columns 2-6) as well as the averaged types for all indices (columns 10-11).
These classifications are
generally equivalent to those determined by direct spectral comparison,
as demonstrated in
Figures~\ref{fig_stdindvsspt_spex} and~\ref{fig_rngindvsspt_spex}
for the SpeX dataset.
Differences between the direct comparison
and index-based classifications are generally less than 0.5-1 subtypes, with typical
deviations of 0.3 subtypes.
Hence, all three techniques yield results that are consistent within the nominal
0.5 subtype uncertainty of the scheme, and any
could be used to classify the spectrum
of a T dwarf.  It is important to stress, however,
that spectral indices are only proxies for the overall spectral
morphology of a source, and direct spectral comparison to the
standards is the most accurate and consistent means of classification
\citep{mor73}.

\section{Discussion}

\subsection{Revised Spectral Types and T Dwarf Properties}

The revised classification scheme for T dwarfs presented here is similar in design
to the schemes proposed by B02 and G02, both in the choice of spectral standards
and spectral indices.  Table~\ref{tab_tcomp} confirms this, comparing
revised spectral types for 61 T dwarfs (based on direct spectral comparison) to
prior classifications made on the B02 and G02 schemes.  Nearly all of
these are consistent within 0.5 subtypes; only seven differ by a full subclass
(including the peculiar T7 Gliese 229B; see below).  Hence, previously identified
trends of absolute brightness \citep{dah02,tin03,vrb04}, color
\citep{me02a,leg02b,kna04} and {\teff} \citep{gol04,vrb04} as a function
of spectral type remain effectively unchanged.

Studies to date have demonstrated that
for subtypes T5 and later, T spectral types are largely correlated with
{\teff} and luminosity \citep{gol04,vrb04}.  This result is broadly consistent with the
observed correlation between MK numerical type and temperature
for main sequence stars. However, the temperature correlation for earlier-type T dwarfs is weak,
with {\teff} roughly constant from L8 to T5
despite significant changes in near infrared
spectral energy distributions
\citep{kir00,me02a,dah02,tin03,gol04,nak04,vrb04}.
The unusual properties of this spectral morphological transition,
as well as the significant
color variations found among similarly-typed late-type T dwarfs
\citep{kna04,me05a}, suggest that further extensions
to the one dimensional
scheme defined here may be required.  However, the clear
variation in spectral morphologies of the standard stars
across this transition
imply that the scheme itself need not be redefined.

\subsection{Extending the T Dwarf Spectral Sequence}

\subsubsection{Dust Effects in Early-type T Dwarf Spectra}

Condensate dust strongly influences the near infrared spectra of
L dwarfs and early-type T dwarfs,
but is largely absent in late-type T dwarf atmospheres \citep{tsu96,all01}.
The depletion of dust from cool brown dwarf
photospheres is now believed to play a greater role in
the spectral transition between
L and T dwarfs than {\teff} \citep{ack01,me02a,tsu03}, although
the mechanism for this depletion remains under considerable debate
\citep{me02c,kna04,tsu05}.
The composition and abundance of condensate dust species,
and the thickness, density and surface distribution of condensate cloud structures \citep{ack01},
are likely to be complex functions of {\teff}, surface gravity, metallicity, rotation
and other parameters \citep{hel01,hel04,lod02,woi03,woi04}; hence, substantial
spectral variations among dust-dominated sources might be expected.
Indeed, there appears to be a decoupling of optical and near infrared spectral
morphologies among late-type L dwarfs and early-type T dwarfs
that could be indicative of such dust/{\teff} effects \citep{cru03,kna04}.
A few of the early-type T dwarfs (e.g., 2MASS 0949$-$1545 and 2MASS 2139+0220; Tinney et al.\ 2005;
Cruz et al.\ in prep.), while having overall spectral morphologies
consistent with T1--T2 dwarfs, exhibit a much broader range of band strengths
(and hence larger scatter in their index subtypes) and $J-K_s$ colors.
While unresolved multiplicity cannot be ruled out for these sources ($\S$~6),
the possible influence of dust-related spectral variations falls beyond the one-dimensional scheme
defined here, and may require an additional classification parameter.

The temporal evolution of dust-sensitive spectral features
must also be characterized, particularly as this directly affects
the stability of the spectral standards.
While a number of studies have measured photometric variability
of up to 0.5~mag in several
L and T dwarfs \citep{bai99,bai01b,tin99,gel02,eno03,bai03,koe03,koe04},
only one T dwarf has been monitored for spectral variations so far
(SDSS~1624+0029; Nakajima et al.\ 2000).
As the photometric variations have been linked
to the surface evolution of condensate dust clouds
\citep{gel02,moh02}, verifying the spectral stability
of the earliest-type T standards, which are most affected by
the photospheric condensates, should be a priority for
future observational studies.

\subsubsection{Gravity Effects in Late-type T Dwarf Spectra}

\begin{figure}
\epsscale{0.8}
\plotone{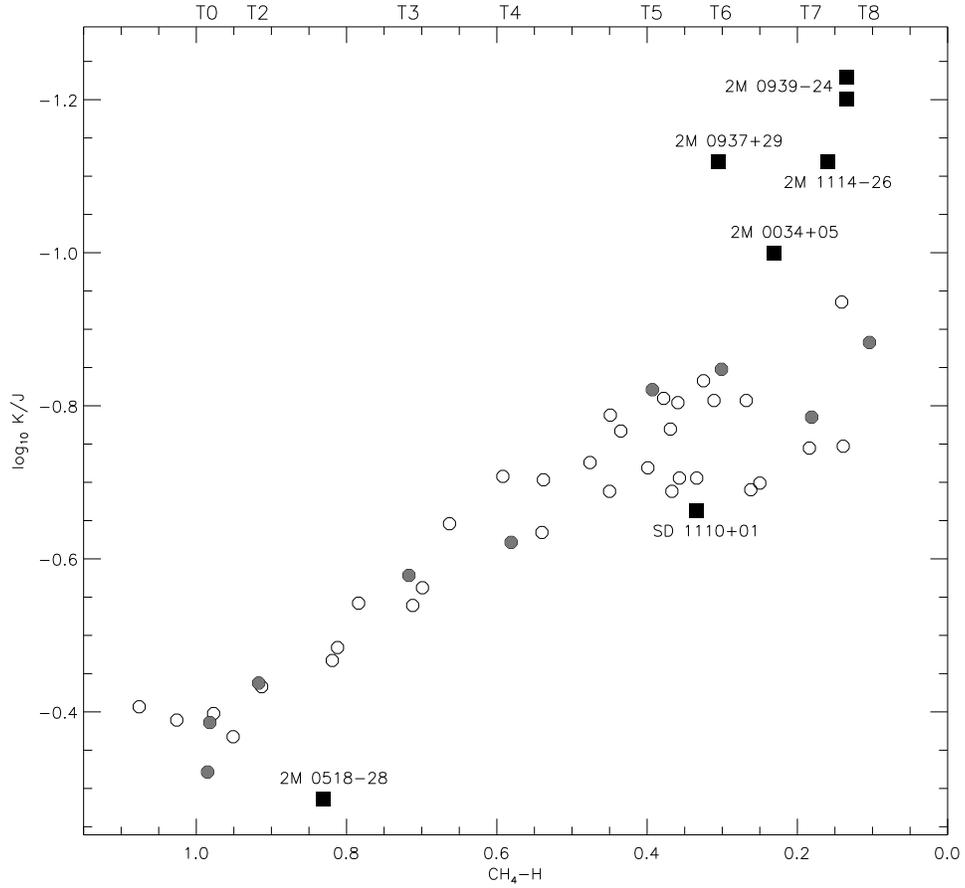}
\caption{Comparison of K/J (on a logarithmic scale) and {\meth}-H indices
as measured for SpeX prism data.  Spectral standards are distinguished by grey circles.
Sources suspected to have unusually high or low
surface gravities or subsolar metallicities
are indicated by black squares and labelled
(2MASS 0518$-$2828 is separately discussed in $\S$~6).
The broad spread in the $K/J$ indices amongst the late-type
T dwarfs suggests that a second classification parameter may be required
to fully describe their spectra.
\label{fig_kjvsch4h}}
\end{figure}

On the other end of the sequence,
a handful of late-type T dwarfs, which exhibit otherwise
normal band strengths, show significant variations
in their $K$-band flux peaks
as compared to the spectral standards.  This phenomenon, cited
extensively in the literature \citep{me02a,me04a,me05a,leg02b,gol04,kna04}, is caused by
variations in collision induced H$_2$ absorption, which is particularly strong in the
atmospheres of the coldest brown dwarfs.
Correlated variations are also seen in the strengths of the 1.25~$\micron$ \ion{K}{1}
lines \citep{kna04} and the 1.05 $\micron$ flux peak \citep{me05a}.
These spectral peculiarities
are likely tied to differences in surface gravities and possibly metallicities
among late-type T dwarfs \citep{sau94,bur02,me05a,kna04},
and can vary significantly for a given set of {\water} and {\meth} band strengths.
This is demonstrated graphically in Figure~\ref{fig_kjvsch4h}, which
compares the K/J and {\meth}-H ratios for the SpeX prism
dataset.\footnote{The SpeX dataset is particularly useful for exploring this phenomenon as the entire
0.8-2.5 $\micron$ range is observed in a single dispersion order, mitigating
flux scaling errors that can affect data obtained in multiple settings.}
While the ratios show reasonable
correlation over much of the T sequence, there is a substantial
increase in the range of K/J values amongst the latest-type T dwarfs.
The decoupling of these ``surface gravity features'' from the standard
classification sequence again
suggests that a second classification parameter is required to
fully describe the spectra.

While dust, gravity and metallicity features may necessitate
extensions to the one dimensional
scheme defined here,
the inclusion of a second (or third or fourth)
classification parameter under the MK Process
requires the identification
of additional spectral standards to map out the new parameter space.
As the number of ``deviant'' T dwarfs is still relatively small,
we leave this task to a future study.

\subsection{To the End of the T Dwarf Class and Beyond}

The classification scheme defined here sufficiently encompasses
the range of spectral morphologies observed for the known
population of T dwarfs.  But what about colder brown dwarfs
that are likely to be identified in the near future?
Assigning new T subtypes for such sources would require that they follow
the same spectral trends as the latest-type T dwarfs,
namely progressively stronger bands of {\water} and {\meth}.
As these bands are already nearly saturated in the spectrum of the T8 standard
2MASS~0415$-$0935, it is unlikely that many more {\em readily distinguishable} T
subtypes will be identified \citep{bur03}.  This is not to say that significantly
cooler brown dwarfs could not be classified as late-type T dwarfs; these
later subtypes may just encompass much broader {\teff} intervals.  Furthermore,
there is no {\em a priori} reason that T subtypes could not extend
to T10 and beyond, as is the case for the M giant sequence.

On the other hand, if the near infrared spectral features of a cold brown
dwarf discovery were significantly and systematically different than those
examined here, a new spectral class, already referred to in the literature
as the Y dwarf class \citep{kir99,bur03,kna04}, would be required.
How will these spectra differ?
Theoretical atmosphere models predict several effects,
including the emergence of NH$_3$ bands at 1.5 and 1.95~$\micron$
below {\teff} $\sim$ 600~K (the 10.5~$\micron$ band has already
been detected in the spectrum of Epsilon Indi Bab; Roellig et al.\ 2004);
the disappearance of the strong pressure-broadened
\ion{Na}{1} and \ion{K}{1}
doublets, and weaker \ion{Cs}{1} lines at 0.85 and 0.89~$\micron$,
below {\teff} $\sim$ 500~K;
the condensation of {\water} vapor around 400--500~K;
and the gradual transition to red near infrared spectral
energy distributions around 300--400~K \citep{bur03}.
Any of these near infrared spectral transitions
would signal a clear break between the T and Y spectral classes.
Such a break could also arise at other wavelengths.
Just as T dwarfs are distinguished from the (originally) optically-classified L
dwarfs by the presence of the near infrared {\meth} bands \citep{kir99,geb02},
so too may Y dwarfs be distinguished by the emergence of distinct
features longward of 2.5~$\micron$, such as the 2.95~$\micron$ NH$_3$ band
or broad H$_2$ features beyond 10~$\micron$ \citep{bur03}.
Until examples of these cold brown dwarfs are actually
identified, however, delineating the end of the T dwarf class is largely
an exercise in speculation.

\section{Individual Sources}

A handful of sources studied here deserve additional attention,
either due to their spectral
peculiarity or significantly revised classification.

\begin{figure}
\epsscale{0.8}
\plotone{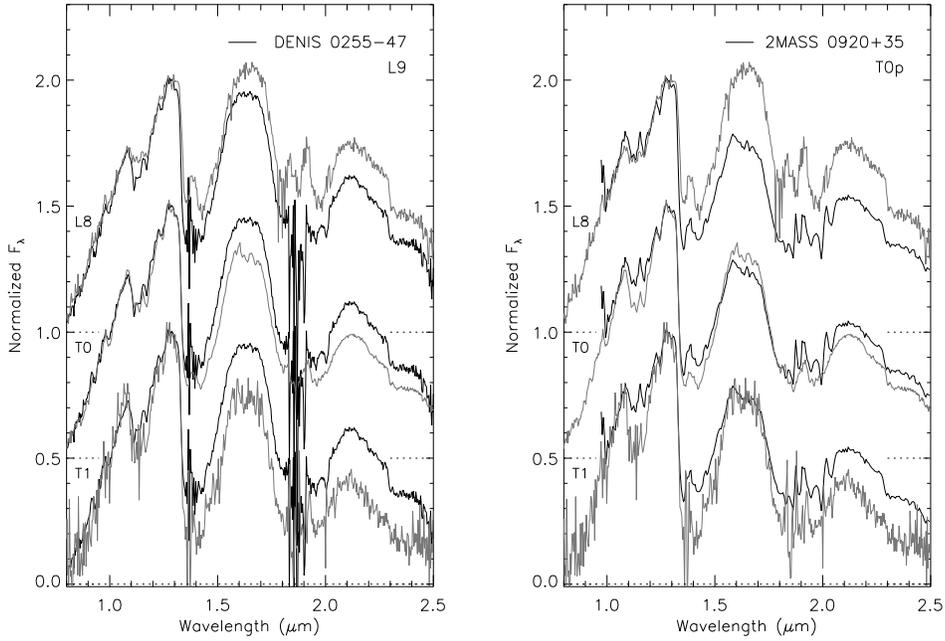}
\caption{{\em Left}: Comparison of SpeX prism data for DENIS 0255$-$4700
(black line) to L8, T0 and T1 spectral standards (grey lines; note that
the alternate standard SDSS 0423$-$0414AB is shown here).  All of the spectra
are normalized at their $J$-band flux peaks and offset by a constant (dotted lines).
While \citet{cus05} have detected {\meth}
absorption at 1.6 $\micron$ in a higher
resolution SpeX spectrum of DENIS 0255$-$4700,
this band is not seen in these low resolution data.
{\em Right:} Similar comparison of NIRC
grism data for 2MASS 0920+3517AB (B02).  In this case {\meth}
absorption at 1.6 $\micron$
is clearly present, although there are discrepancies
in band strengths; e.g., the {\water}/{\meth} band at 1.1 $\micron$
as compared to the {\meth} band at 1.6 $\micron$.
This source is therefore classified T0p in the near infrared, much
later than its L6.5 optical classification \citep{kir00}.
\label{fig_0255and0920}}
\end{figure}

\noindent {\it DENIS 0255$-$4700 (L9):} Identified by \citet{mrt99} in the Deep Near Infrared Survey
of the Southern Sky (DENIS; Epchtein et al.\ 1997), this relatively bright L dwarf
(2MASS $J$=13.25$\pm$0.03) is classified L8 in the optical (Kirkpatrick et al.\ in prep.).
\citet{cus05} have detected
weak {\meth} absorption at 1.6 and 2.2 $\micron$ in moderate resolution SpeX data,
and suggest that this source could be classified as a T dwarf based on the definition
put forth by G02.  However, it is important to note that the analogous definition adopted here
requires the detection of $H$-band {\meth} absorption
at low spectral resolutions.  Examination of SpeX prism
data (Burgasser et al.\ in prep.; Figure~\ref{fig_0255and0920}) reveals
no distinct $H$-band {\meth} absorption feature (likely due to its weakness)
and an overall spectral energy distribution consistent with a late-type L dwarf.
Therefore, according to the scheme defined here,
DENIS 0255$-$4700 is classified $\sim$L9 in the near infrared,
of slightly earlier type than a T dwarf.

\noindent {\it SDSS 1104+5548 and SDSS 2047$-$0718 (T0:):} Classified L9.5$\pm$1.5 and L9.5$\pm$1.0
by G02, respectively, these faint sources are reclassified T0{:} based on their
similarity to the T0 standard SDSS 1207+0244.  Weak {\meth} absorption appears to be present at
both 1.6 and 2.2 $\micron$, although the low S/N spectra of both sources make these assignments
uncertain.

\noindent {\it 2MASS 0518$-$2828 (T1p):}  The spectrum of this source is clearly peculiar
(Figure~\ref{fig_spexviscomp}), with $J$- and $H$-band spectral
features consistent with a subtype later than T2, but a
$K$-band spectrum and red $J-K$ color consistent with a late-type L dwarf/early-type T dwarf.
\citet{cru04}, who identified the source in the 2MASS database, have proposed that it
is an unresolved binary composed of a
mid/late-type L dwarf ($\sim$L6) and early/mid-type T dwarf ($\sim$T4).
Confirmation of this hypothesis is forthcoming (Cruz 2005, priv.\ comm.).

\noindent {\it 2MASS 0920+3517AB (T0p):} This source was identified by \citet{kir00} in
the 2MASS database and is classified L6.5 in the optical.
Spectroscopic observations by \citet{nak01} reveal the
presence of {\meth} absorption at 1.6 and 2.2 $\micron$,
which are verified, but previously unrecognized, in low resolution NIRC
grism data (B02; Figure~\ref{fig_0255and0920}).  However,
there are some spectral discrepancies; the 1.1 $\micron$ {\water}/{\meth} band
is shallower than those observed in early-type T dwarfs, while
the 2.2 $\micron$ {\meth} band is fairly weak compared
to the 1.6 $\micron$ band.  Like 2MASS 0518$-$2828, these
features may arise from a composite spectrum of a late-type L dwarf
and early-type T dwarf pair. 2MASS 0920+3517AB is marginally
resolved as a closely-separated ($\sim$0$\farcs$075)
binary in {\em HST} observations \citep{rei01}.\footnote{\citet{rei01}
also speculate that the 2MASS 0850$-$1057AB system, the
L6 spectral standard on the \citet{kir99} optical scheme,
may harbor a T dwarf secondary.} \citet{bou03} determine
an F814W ($\lambda_c$ = 0.83 $\micron$) flux ratio of 0.88$\pm$0.11 mag,
implying $M_I \approx 19$ for 2MASS 0920+3517B,
comparable to the $I$-band absolute magnitudes of
late-type L dwarfs through mid-type T
dwarfs \citep{dah02}.  A composite L/T spectrum might also explain
the large discrepancy between the optical spectral type and near infrared
spectral morphology of this source.

\noindent {\it Gliese 337CD (T0):} Identified by \citet{wil01} as a widely separated
(43$\arcsec$ $\approx$ 880 AU)
common proper motion companion to the G8V+K1V binary Gliese 337AB, and itself
resolved as a nearly equal brightness binary at $K$-band \citep{me05b}, this source
is optically classified L8 on the \citet{kir99} scheme.  Its near infrared
spectrum \citep{mcl03} exhibits distinct signatures of {\meth} absorption
at both 1.6 and 2.2 $\micron$ that are similar to
features in the T0 standard SDSS 1207+0244.  The composite spectrum
is therefore reclassified T0 here.
The small difference in absolute $K$-band
magnitudes from L8--T4 \citep{tin03,gol04,vrb04} suggests
that this system could also be composed of a late-type L dwarf and
early-type T dwarf pair.

\noindent {\it 2MASS 0937+2931 (T6p):}  The prototype example of photospheric
pressure effects
in T dwarf spectra ($\S$~5.2.2), this peculiar T6 has been discussed extensively in the
literature \citep{me02a,me03d,gol04,kna04,vrb04,me05a}.  The current scheme
retains the T6 classification of B02, but spectral peculiarities are
clearly evident in both low resolution (enhanced 1.05 $\micron$ peak, suppressed
$K$-band peak) and moderate resolution (weak 1.25 $\micron$ \ion{K}{1} lines)
data. 2MASS~0034+0523, 2MASS~0939$-$2448 and 2MASS~1114$-$2618 show similar, albeit
less pronounced, spectral deviations \citep{me04a,me05a}; while SDSS~1110+0116
exhibits enhanced $K$-band flux, suggesting that it is a low gravity source
\citep{kna04}.  The peculiar status of 2MASS~0937+2931 may
therefore simply reflect the shortcomings
of the one-dimensional classification scheme defined here.

\noindent {\it Gliese 229B (T7p):} One of the first brown dwarfs
to be identified \citep{nak95,opp95},
this prototype T dwarf
nevertheless fails to fit cleanly within the sequence of spectral standards.
As shown in Figure~\ref{fig_cgs4viscomp}, the CGS4 spectrum
of Gliese 229B \citep{geb96} exhibits strong {\meth} absorption
at 1.3 and 1.6 $\micron$ consistent with a T7 spectral type; but weaker
{\water} and {\meth} bands at 1.1, 1.4, 1.9 and 2.2 $\micron$, consistent with
type T5--T6.  Low resolution NIRC grism data \citep{opp98} show similar deviations.
Because of its early discovery, there have been several detailed studies of Gliese 229B
suggesting that it may have atypical physical properties.  \citet{gri98} and \citet{gri00} have
proposed that the metallicity of Gliese 229B may be 0.3-0.5 solar, based on
spectral model fits to {\water} and \ion{Cs}{1} features in its optical spectrum.
Using spectral model fits for the M1V primary,
\citet{leg02a} have also concluded that this system is metal-poor
and possibly young as well.
Both Gliese 229B and 2MASS~0937+2931 are therefore
important sources for studying the role of surface gravity and metallicity
on the emergent spectra of the coldest known brown dwarfs.

\section{Summary}

We have defined a revised near infrared classification scheme for
T dwarfs that unifies and supersedes original schemes proposed by
B02 and G02.  Nine primary spectral standards and five alternate standards
define the scheme, and classifications for other
sources may be made by the direct comparison of equivalent spectral data.
Two alternate methods of classification
using spectral indices have been described and shown to
have no significant differences with direct spectral comparison.
The revised classifications are also consistent with the prior
schemes, implying that existing analyses of spectral type trends remain intact.
Nonetheless, we point out that future extensions to the one dimensional
classification scheme presented here may be needed to incorporate additional
spectral variations.  These extensions may be made when
more examples of these objects are identified and characterized.

In the Appendix, we provide a compendium of all T dwarfs
with published near infrared spectra
at the time this article was written,
listing classifications on the revised scheme.
With 69 systems comprising this census, the study of cold brown dwarfs
has crossed a threshold where detailed comparative analyses, population statistics
and the discovery and characterization of physically unique sources can be made.
As forthcoming deep and wide-field sky surveys
(e.g., UKIDSS, Warren 2002) uncover
many more of our T dwarf neighbors,
a unified framework for classifying
these objects is clearly essential.

\acknowledgments

The authors give special thanks to
K.\ Cruz, J.\ G.\ Cuby, M.\ Cushing, M.\ Liu, M.\ McCaughrean,
I.\ S.\ McLean, T.\ Nakajima, B.\ Oppenheimer, R.-D.\ Scholz and M.\ R.\ Zapatero Osorio
for providing published and unpublished spectral data for this work,
and to K.\ Chiu and X.\ Fan for assistance with the retrieval of SDSS photometry.
A.\ J.\ B.\ also thanks C.\ Corbally and R.\ Gray for useful insights on the MK Process.
We thank our anonymous referee for her/his helpful critique of the
original manuscript.
A.\ J.\ B.\ acknowledges support from NASA through the SIRTF Fellowship Program.
T.\ R.\ G.\ acknowledges support by the Gemini Observatory, which is
operated by the Association of Universities for Research in
Astronomy, Inc., on behalf of the international Gemini partnership
of Argentina, Australia, Brazil, Canada, Chile, the United Kingdom
and the United States of America.
Some data were obtained through the UKIRT Service Programme. UKIRT is operated by the
Joint Astronomy Centre on behalf of the UK Particle Physics and Astronomy Research Council.
This publication makes use of data from the Two Micron All Sky Survey, which is a
joint project of the University of Massachusetts and the Infrared
Processing and Analysis Center, and funded by the National
Aeronautics and Space Administration and the National Science
Foundation. 2MASS data were obtained from the NASA/IPAC Infrared
Science Archive, which is operated by the Jet Propulsion
Laboratory, California Institute of Technology, under contract
with the National Aeronautics and Space Administration.
Funding for the Sloan Digital Sky Survey (SDSS) has been provided by
the Alfred P. Sloan Foundation, the Participating Institutions, the
National Aeronautics and Space Administration, the National Science
Foundation, the U.S. Department of Energy, the Japanese Monbukagakusho,
and the Max Planck Society. The SDSS Web site is \url{http://www.sdss.org/}.
The SDSS is managed by the Astrophysical Research Consortium (ARC)
for the Participating Institutions. The Participating Institutions
are The University of Chicago, Fermilab, the Institute for Advanced Study,
the Japan Participation Group, The Johns Hopkins University, the Korean
Scientist Group, Los Alamos National Laboratory, the Max-Planck-Institute
for Astronomy (MPIA), the Max-Planck-Institute for Astrophysics (MPA),
New Mexico State University, University of Pittsburgh, University of Portsmouth,
Princeton University, the United States Naval Observatory, and the
University of Washington.

\appendix

\section{T Dwarf Compendium}

Table~\ref{tab_allt} provides a compendium of all T dwarfs
with published near infrared spectra
at the writing of this article,
and a sample of their empirical properties.
This list includes 69 systems, including nine binaries and four companions
to main sequence stars.
Near infrared spectral types on the revised scheme are provided in column 2.
Equinox J2000 coordinates, listed in columns 3 and 4, are
primarily from the 2MASS All Sky Point Source
Catalog \citep{cut03} with the exceptions of
IFA 0226+0051, S Ori 70, SDSS 0837$-$0000,
SDSS 1104+5548, SDSS 1157+0611, NTTDF 1205$-$0744, SDSS 1632+4150 and SDSS 2047-0718
for which positional data were obtained from the SDSS catalog or
the discovery reference.  Coordinate epochs are given in column 5.
SDSS $z$-band magnitudes on the AB asinh system \citep{fuk96,lup99}
from the SDSS Data Release 4 \citep{ade05} are listed in column 6.
2MASS $JHK_s$ photometry from the All Sky Catalog is listed
in columns 7--9; for those sources undetected or marginally
detected by 2MASS, alternate CIT or MKO\footnote{Note that photometry on the
MKO system can be very different from photometry on the 2MASS or CIT systems.
Values measured for T dwarfs in the
former system can be found in \citet{leg02b} and \citet{kna04},
and transformations
between the MKO and 2MASS systems for late-type dwarfs
are given in \citet{ste04}.} $JHK$ photometry is provided.
Parallax ($\pi$; column 10) and proper motion ($\mu$, $\phi$; columns 11 and 12)
measurements from \citet{dah02,tin03}; and \citet{vrb04} are listed
for field sources; and from HIPPARCOS \citep{esa97} for T dwarf companions
to main sequence stars.  Additional proper motion measurements
from \citet{me03a,me04a,me03e,ell05}; and \citet{tin05} are also listed.
Relevant publications, including discovery citations, are provided in column 13.
An electronic version of this table will be made available through the journal.

\clearpage

\begin{deluxetable}{lllcll}
\tabletypesize{\scriptsize}
\tablecaption{Spectral Datasets Examined for T Dwarf Classification. \label{tab_speclib}}
\tablewidth{0pt}
\tablehead{
\colhead{} &
\colhead{} &
\colhead{Spectral} &
\colhead{Number of} &
\colhead{Number of} &
\colhead{} \\
\colhead{Instrument} &
\colhead{\ldl} &
\colhead{Coverage} &
\colhead{Orders} &
\colhead{Sources} &
\colhead{References} \\
}
\startdata
CTIO 4m OSIRIS & 1200 & 1.1--2.4 $\micron$ & 4 & 20 T, 0 L & 1,2 \\
{\bf IRTF SpeX (prism)}\tablenotemark{a} & 150  & 0.8--2.5 & 1 & 43 T, 2 L & 3,4,5,6 \\
IRTF SpeX (SXD) & 1200  & 0.9--2.5 $\micron$  & 4--6 & 8 T, 1 L & 3,7  \\
Keck NIRC & 100 & 0.9--2.5 $\micron$  & 1--2 & 17 T, 0 L  & 1,8,9 \\
Keck NIRSPEC & 2000 & Various & 1--6 & 17 T, 2 L & 10,11  \\
NTT ISAAC+SOFI & 50 & 0.95--2.5 $\micron$ & 2 & 1 T, 0 L & 12  \\
Palomar 1.5m CORMASS & 300 & 0.8--2.4 $\micron$ & 4 & 1 T, 0 L & 13  \\
Subaru CISCO & 500 & 0.9--2.5 $\micron$ & 3 & 2 T, 0 L & 14  \\
\enddata
\tablenotetext{a}{Primary classification spectral sample.}
\tablerefs{(1) B02; (2) \citet{me03e}; (3) \citet{me04a}; (4) \citet{cru04};
(5) \citet{me05a}; (6) Burgasser et al., in prep.;
(7) \citet{cus05}; (8) \citet{opp95}; (9) \citet{zap02}; (10) \citet{mcl03};
(11) \citet{liu02}; (12) \citet{cub99}; (13) \citet{me00c};
(14) \citet{nak04}; (15) \citet{nak01};
(16) \citet{geb96};
(17) \citet{str99}; (18) \citet{tsv00}; (19) \citet{leg00}; (20) \citet{geb01};
(21) G02; (22) \citet{kna04}; (23) \citet{mcc04}}
\tablecomments{OSIRIS (Ohio State InfraRed Imager/Spectrometer): \citet{dep93};
SpeX: \citet{ray03};
NIRC (Near InfraRed Camera): \citet{mat94};
NIRSPEC (Near-IR Spectrometer): \citet{mcl98,mcl00b};
ISAAC (Infrared Spectrometer and Array Camera): \citet{mor97};
SOFI (Son of Isaac): \citet{mor98};
CORMASS (Cornell Massachusetts Slit Spectrograph): \citet{wil01a};
CISCO (Cooled Infrared Spectrograph and Camera for OHS): \citet{mot02};
IRCS (Infrared Camera and Spectograph): \citet{kob00};
CGS4 (Cooled Grating Spectrometer 4): \citet{wri93};
NAOS/CONICA: \citet{len98}
}
\end{deluxetable}

\clearpage

\begin{deluxetable}{lccccccl}
\tabletypesize{\scriptsize}
\tablecaption{T Dwarf Spectral Standards. \label{tab_standards}}
\tablewidth{0pt}
\tablehead{
 &
\colhead{NIR} &
\multicolumn{4}{c}{MKO Photometry\tablenotemark{a}} & &
\colhead{Spectral} \\
\cline{3-6}
\colhead{Name} &
\colhead{SpT} &
\colhead{$J$} &
\colhead{$J-H$} &
\colhead{$H-K$} &
\colhead{$K-L^{\prime}$} &
\colhead{d (pc)\tablenotemark{b}} &
\colhead{Sample\tablenotemark{c}} \\
\colhead{(1)} &
\colhead{(2)} &
\colhead{(3)} &
\colhead{(4)} &
\colhead{(5)} &
\colhead{(6)} &
\colhead{(7)} &
\colhead{(8)} \\
}
\startdata
\cline{1-8}
\multicolumn{8}{c}{Primary Standards} \\
\cline{1-8}
SDSS J120747.17+024424.8 & T0 &  15.38$\pm$0.03 &  0.75$\pm$0.04 & 0.47$\pm$0.04 & 1.54$\pm$0.06 & \nodata & 1 \\
SDSS J083717.21$-$000018.0 & T1 &  16.90$\pm$0.05 & 0.69$\pm$0.07  &  0.23$\pm$0.07  & \nodata & 29$\pm$12 & 1,2,3 \\
SDSS J125453.90$-$012247.4 & T2 &  14.66$\pm$0.03 &  0.53$\pm$0.04  &  0.29$\pm$0.04  &  1.59$\pm$0.06 & 13.7$\pm$0.4 & 1,2,4,5,6,7 \\
2MASS J12095613$-$1004008 & T3 & 15.55$\pm$0.03 & 0.31$\pm$0.04 & 0.07$\pm$0.04 & \nodata & \nodata & 1,2 \\
2MASS J22541892+3123498 & T4 &  15.01$\pm$0.03  & 0.06$\pm$0.04  &  $-$0.08$\pm$0.04  &  1.79$\pm$0.06 & \nodata & 1,2,3,5,7 \\
2MASS J15031961+2525196 & T5 &  13.55$\pm$0.03 &   $-0.35$$\pm$0.04  &  $-0.09$$\pm$0.04 &  2.08$\pm$0.06 & \nodata & 2,4 \\
SDSS J162414.37+002915.6 & T6 &  15.20$\pm$0.03  & $-0.28$$\pm$0.04 &  $-0.13$$\pm$0.04 &  2.01$\pm$0.05 & 11.00$\pm$0.15 & 1,2,5,7 \\
\enddata
\tablenotetext{a}{MKO $JHKL^{\prime}$ photometry from \citet{str99,leg00,leg02b}; G02; \citet{gol04}; and \citet{kna04}.}
\tablenotetext{b}{Parallax measurements from \citet{dah02,tin03}; and \citet{vrb04}.}
\tablenotetext{c}{Spectral samples as follows: (1) UKIRT CGS4; (2) IRTF SpeX;
(3) Keck NIRC; (4) IRTF SpeX cross-dispersed mode;
(5) Keck NIRSPEC; (6) Subaru CISCO; (7) CTIO OSIRIS.  See Table~\ref{tab_speclib}.}
\end{deluxetable}

\begin{deluxetable}{lccl}
\tabletypesize{\scriptsize}
\tablecaption{Definitions of Near Infrared Spectral Indices. \label{tab_indices}}
\tablewidth{0pt}
\tablehead{
\colhead{} &
\colhead{Numerator} &
\colhead{Denominator} &
\colhead{}  \\
\colhead{Index} &
\colhead{Range\tablenotemark{a}} &
\colhead{Range\tablenotemark{a}} &
\colhead{Feature} \\
\colhead{(1)} &
\colhead{(2)} &
\colhead{(3)} &
\colhead{(4)} \\
}
\startdata
{\bf H$_2$O-J} & 1.140--1.165 & 1.260--1.285 & 1.15 $\micron$ H$_2$O \\
{\bf CH$_4$-J} & 1.315--1.340 & 1.260--1.285 & 1.32 $\micron$ CH$_4$ \\
{\bf H$_2$O-H} & 1.480--1.520 & 1.560--1.600 & 1.4 $\micron$ H$_2$O \\
{\bf CH$_4$-H} & 1.635--1.675 & 1.560--1.600 & 1.65 $\micron$ CH$_4$ \\
H$_2$O-K & 1.975--1.995 & 2.080--2.100 & 1.9 $\micron$ H$_2$O \\
{\bf CH$_4$-K} & 2.215--2.255 & 2.080--2.120 & 2.2 $\micron$ CH$_4$ \\
K/J & 2.060--2.100 & 1.250--1.290 & $J-K$ color \\
\enddata
\tablenotetext{a}{Wavelength range (in microns) over which flux density ($f_{\lambda}$) is integrated.}
\tablecomments{Primary classification indices are listed in boldface type.}
\end{deluxetable}

\clearpage

\begin{deluxetable}{llccccc}
\tabletypesize{\scriptsize}
\tablecaption{Near Infrared Spectral Indices for T Dwarf Spectral Standards. \label{tab_indstandard}}
\tablewidth{0pt}
\tablehead{
\colhead{Name} &
\colhead{NIR SpT} &
\colhead{H$_2$O-J} &
\colhead{CH$_4$-J} &
\colhead{H$_2$O-H} &
\colhead{CH$_4$-H} &
\colhead{CH$_4$-K} \\
\colhead{(1)} &
\colhead{(2)} &
\colhead{(3)} &
\colhead{(4)} &
\colhead{(5)} &
\colhead{(6)} &
\colhead{(7)}  \\
}
\startdata
\cline{1-7}
\multicolumn{7}{c}{SpeX Prism Data} \\
\cline{1-7}
2MASS 1632+1904 & L8\tablenotemark{a} & 0.706  & 0.735  & 0.705  & 1.077 & 0.881  \\
SDSS 0423$-$0414\tablenotemark{b} & T0 & 0.630  & 0.644  & 0.621  & 0.985  & 0.820 \\
SDSS 0837$-$0000 & T1 & 0.572  & 0.614  & 0.539  & 0.982  & 0.693 \\
SDSS 1254$-$0122 & T2 & 0.474  & 0.583  & 0.474  & 0.917  & 0.585 \\
2MASS 1209$-$1004 & T3 & 0.413  & 0.516  & 0.453  & 0.717  & 0.496 \\
2MASS 2254+3123 & T4 & 0.369  & 0.506  & 0.389  & 0.581  & 0.305 \\
2MASS 1503+2525 & T5 & 0.240  & 0.356  & 0.345  & 0.393  & 0.200 \\
\enddata
\tablenotetext{a}{{Late} L dwarf comparison source.}
\tablenotetext{b}{Alternate standard.}
\tablenotetext{c}{With the exception of SpeX cross-dispersed data for 2MASS 1503+2525 \citep{me04a}.}
\end{deluxetable}

\begin{deluxetable}{cccccc}
\tabletypesize{\scriptsize}
\tablecaption{Numerical Ranges of Near Infrared Spectral Indices for T Dwarf Subtypes. \label{tab_indrng}}
\tablewidth{0pt}
\tablehead{
\colhead{NIR SpT} &
\colhead{H$_2$O-J} &
\colhead{CH$_4$-J} &
\colhead{H$_2$O-H} &
\colhead{CH$_4$-H} &
\colhead{CH$_4$-K} \\
\colhead{(1)} &
\colhead{(2)} &
\colhead{(3)} &
\colhead{(4)} &
\colhead{(5)} &
\colhead{(6)} \\
}
\startdata
T0 & \nodata & 0.73-0.78 & 0.60--0.66 & 0.97--1.00 & 0.75--0.85 \\
T1 & $>$ 0.55 & 0.67-0.73 & 0.53--0.60 & 0.92--0.97 & 0.63--0.75 \\
T2 & 0.45--0.55 & 0.58-0.67 & 0.46--0.53 & 0.80--0.92 & 0.55--0.63 \\
T3 & 0.38--0.45 & 0.52-0.58 & 0.43--0.46 & 0.60--0.80 & 0.35--0.55 \\
T4 & 0.32--0.38 & 0.45-0.52 & 0.37--0.43 & 0.48--0.60 & 0.24--0.35 \\
T5 & 0.18--0.32 & 0.36-0.45 & 0.32--0.37 & 0.36--0.48 & 0.18--0.24 \\
T6 & 0.13--0.18 & 0.28-0.36 & 0.26--0.32 & 0.25--0.36 & 0.13--0.18 \\
T7 & 0.07--0.13 & 0.21-0.28 & 0.20--0.26 & 0.15--0.25 & $<$ 0.13 \\
T8 & 0.02--0.07 & 0.15-0.21 & 0.14--0.20 & 0.07--0.15 & \nodata \\
\enddata
\end{deluxetable}

\clearpage

\begin{deluxetable}{llllllcc|lll}
\tabletypesize{\scriptsize}
\tablecaption{Spectral Indices and Classifications for SpeX Prism Spectra. \label{tab_spexclass}}
\tablewidth{0pt}
\rotate
\tablehead{
\colhead{} &
\multicolumn{7}{c}{} &
\multicolumn{3}{c}{Derived NIR SpT} \\
\cline{9-11}
\colhead{Name} &
\multicolumn{1}{c}{H$_2$O-J} &
\multicolumn{1}{c}{CH$_4$-J} &
\multicolumn{1}{c}{H$_2$O-H} &
\multicolumn{1}{c}{CH$_4$-H} &
\multicolumn{1}{c}{CH$_4$-K} &
\multicolumn{1}{c}{H$_2$O-K} &
\multicolumn{1}{c}{K/J} &
\colhead{Direct} &
\colhead{Ind 1} &
\colhead{Ind 2} \\
\colhead{(1)} &
\colhead{(2)} &
\colhead{(3)} &
\colhead{(4)} &
\colhead{(5)} &
\colhead{(6)} &
\colhead{(7)} &
\colhead{(8)} &
\colhead{(9)} &
\colhead{(10)} &
\colhead{(11)} \\
}
\startdata
2M 1632+1904 & 0.706 (L8.0/--) & 0.735 (L8.0/T0.0) & 0.705 (L8.0/--) & 1.077 (--/--) & 0.881 (L8.0/--) & 0.696 & 0.743 & {\bf L8.0} & L8.0 & \nodata  \\
DE 0255-4700 & 0.671 (L8.0/--) & 0.633 (T0.5/T2.0) & 0.688 (L8.0/--) & 1.030 (--/--) & 0.835 (T0.0/T0.0) & 0.691 & 0.603 & {\bf L9.0} & L9.0: & \nodata  \\
SD 0423-0414 & 0.630 (T0.0/--) & 0.644 (T0.0/T2.0) & 0.621 (T0.0/T0.0) & 0.985 (T0.0/T0.0) & 0.820 (T0.0/T0.0) & 0.678 & 0.477 & {\bf T0.0} & T0.0 & T0.5:  \\
SD 0151+1244 & 0.654 (L9.5/--) & 0.621 (T1.0/T2.0) & 0.633 (T0.0/T0.0) & 0.951 (T1.5/T1.0) & 0.689 (T1.0/T1.0) & 0.614 & 0.429 & {\bf T0.5} & T0.5 & T1.0  \\
SD 0837-0000 & 0.573 (T1.0/--) & 0.616 (T1.0/T2.0) & 0.560 (T0.5/T1.0) & 0.977 (T1.0/T0.5) & 0.754 (T0.5/T0.5) & 0.608 & 0.400 & {\bf T1.0} & T1.0 & T1.0  \\
 & 0.572 (T1.0/--) & 0.614 (T1.0/T2.0) & 0.539 (T1.0/T1.5) & 0.982 (T1.0/T0.0) & 0.693 (T1.0/T1.0) & 0.594 & 0.411 & {\bf T1.0} & T1.0 & T1.0  \\
2M 2139+0220 & 0.423 (T3.0/T3.0) & 0.503 (T4.0/T4.0) & 0.475 (T2.0/T2.5) & 1.026 (--/--) & 0.685 (T1.0/T1.0) & 0.582 & 0.408 & {\bf T1.5} & T2.5: & T2.5:  \\
 & 0.410 (T3.0/T3.0) & 0.634 (T0.5/T2.0) & 0.455 (T3.0/T2.5) & 1.076 (--/--) & 0.687 (T1.0/T1.0) & 0.561 & 0.392 & {\bf T1.5} & T2.0: & T2.0  \\
\enddata
\tablecomments{Index subtypes in parentheses determined by direct comparison to spectral standard
ratios (Table~\ref{tab_indstandard}) or spectral ratio ranges (Table~\ref{tab_indrng}); see $\S$~4.2.2.
Uncertain classifications are indicated by ``:''; peculiar sources are indicated by ``p''.}
\end{deluxetable}

\clearpage

\begin{deluxetable}{llllllcc|lll}
\tabletypesize{\scriptsize}
\tablecaption{Spectral Indices and Classifications for CGS4 Spectra. \label{tab_cgs4class}}
\tablewidth{0pt}
\rotate
\tablehead{
\colhead{} &
\multicolumn{7}{c}{} &
\multicolumn{3}{c}{Derived NIR SpT} \\
\cline{9-11}
\colhead{Name} &
\multicolumn{1}{c}{H$_2$O-J} &
\multicolumn{1}{c}{CH$_4$-J} &
\multicolumn{1}{c}{H$_2$O-H} &
\multicolumn{1}{c}{CH$_4$-H} &
\multicolumn{1}{c}{CH$_4$-K} &
\multicolumn{1}{c}{H$_2$O-K} &
\multicolumn{1}{c}{K/J} &
\colhead{Direct} &
\colhead{Ind 1} &
\colhead{Ind 2} \\
\colhead{(1)} &
\colhead{(2)} &
\colhead{(3)} &
\colhead{(4)} &
\colhead{(5)} &
\colhead{(6)} &
\colhead{(7)} &
\colhead{(8)} &
\colhead{(9)} &
\colhead{(10)} &
\colhead{(11)} \\
}
\startdata
2M 1632+1904 & 0.701 (L8.0/--) & 0.722 (L8.0/T0.5) & 0.705 (L8.0/--) & 1.036 (--/--) & 0.888 (L8.0/--) & 0.700 & 0.769 & {\bf L8.0} & L8.0 & \nodata  \\
2M 0310+1648 & 0.675 (L9.0/--) & 0.736 (L9.0/T0.5) & 0.645 (L9.0/T0.0) & 1.064 (--/--) & 0.786 (L9.0/T0.0) & 0.667 & 0.683 & {\bf L9.0} & L9.0 & \nodata  \\
\enddata
\tablecomments{Index subtypes in parentheses determined by direct comparison to spectral standard
ratios (Table~\ref{tab_indstandard}) or spectral ratio ranges (Table~\ref{tab_indrng}); see $\S$~4.2.2.
Uncertain classifications are indicated by ``:''; peculiar sources are indicated by ``p''.}
\end{deluxetable}

\clearpage

\begin{deluxetable}{llllllcc|lll}
\tabletypesize{\scriptsize}
\tablecaption{Spectral Indices and Classifications for NIRC Grism Spectra. \label{tab_nircclass}}
\tablewidth{0pt}
\rotate
\tablehead{
\colhead{} &
\multicolumn{7}{c}{} &
\multicolumn{3}{c}{Derived NIR SpT} \\
\cline{9-11}
\colhead{Name} &
\multicolumn{1}{c}{H$_2$O-J} &
\multicolumn{1}{c}{CH$_4$-J} &
\multicolumn{1}{c}{H$_2$O-H} &
\multicolumn{1}{c}{CH$_4$-H} &
\multicolumn{1}{c}{CH$_4$-K} &
\multicolumn{1}{c}{H$_2$O-K} &
\multicolumn{1}{c}{K/J} &
\colhead{Direct} &
\colhead{Ind 1} &
\colhead{Ind 2} \\
\colhead{(1)} &
\colhead{(2)} &
\colhead{(3)} &
\colhead{(4)} &
\colhead{(5)} &
\colhead{(6)} &
\colhead{(7)} &
\colhead{(8)} &
\colhead{(9)} &
\colhead{(10)} &
\colhead{(11)} \\
}
\startdata
2M 0920+3517 & 0.761 (L8.0/--) & 0.635 (T0.5/T2.0) & 0.716 (L8.0/--) & 0.964 (T1.5/T0.5) & 0.860 (L8.0/--) & 0.653 & 0.538 & {\bf T0.0p} & L9.0: & T1.5:  \\
2M 2254+3123 & 0.453 (T2.5/T2.5) & 0.366 (T5.0/T5.5) & 0.425 (T3.5/T3.5) & 0.558 (T4.0/T4.0) & 0.266 (T4.5/T4.5) & 0.549 & 0.191 & {\bf T4.0} & T4.0 & T4.0:  \\
2M 0559-1404 & 0.366 (T4.0/T3.5) & 0.405 (T4.5/T5.0) & 0.375 (T4.5/T4.5) & 0.491 (T4.5/T4.5) & 0.228 (T4.5/T4.5) & 0.445 & 0.252 & {\bf T4.5} & T4.5 & T4.5  \\
 & 0.322 (T4.5/T4.5) & 0.437 (T4.5/T4.5) & 0.347 (T5.0/T5.0) & 0.509 (T4.5/T4.5) & 0.243 (T4.5/T4.5) & 0.524 & 0.275 & {\bf T4.5} & T4.5 & T4.5  \\
2M 2339+1352 & 0.303 (T4.5/T4.5) & 0.232 (T7.0/T7.0) & 0.317 (T5.5/T5.5) & 0.424 (T5.0/T5.0) & 0.200 (T5.0/T5.0) & 0.511 & 0.169 & {\bf T5.0} & T5.5 & T5.5  \\
2M 2356-1553 & 0.250 (T5.0/T5.0) & 0.403 (T4.5/T5.0) & 0.299 (T5.5/T6.0) & 0.352 (T5.5/T5.5) & 0.160 (T6.0/T6.0) & 0.495 & 0.228 & {\bf T5.5} & T5.5 & T5.5  \\
2M 0243-2453 & 0.161 (T6.0/T6.0) & 0.267 (T7.0/T6.5) & 0.278 (T6.0/T6.0) & 0.327 (T5.5/T6.0) & 0.158 (T6.0/T6.0) & 0.494 & 0.198 & {\bf T6.0} & T6.0 & T6.0  \\
2M 0937+2931 & 0.177 (T5.5/T5.5) & 0.290 (T6.5/T6.5) & 0.281 (T6.0/T6.0) & 0.324 (T5.5/T6.0) & 0.236 (T4.5/T4.5) & 0.634 & 0.121 & {\bf T6.0p} & T5.5 & T5.5  \\
\enddata
\tablecomments{Index subtypes in parentheses determined by direct comparison to spectral standard
ratios (Table~\ref{tab_indstandard}) or spectral ratio ranges (Table~\ref{tab_indrng}); see $\S$~4.2.2.
Uncertain classifications are indicated by ``:''; peculiar sources are indicated by ``p''.}
\end{deluxetable}

\begin{deluxetable}{llllllcc|lll}
\tabletypesize{\scriptsize}
\tablecaption{Spectral Indices and Classifications for SpeX Cross Dispersed Spectra. \label{tab_spexhrclass}}
\tablewidth{0pt}
\rotate
\tablehead{
\colhead{} &
\multicolumn{7}{c}{} &
\multicolumn{3}{c}{Derived NIR SpT} \\
\cline{9-11}
\colhead{Name} &
\multicolumn{1}{c}{H$_2$O-J} &
\multicolumn{1}{c}{CH$_4$-J} &
\multicolumn{1}{c}{H$_2$O-H} &
\multicolumn{1}{c}{CH$_4$-H} &
\multicolumn{1}{c}{CH$_4$-K} &
\multicolumn{1}{c}{H$_2$O-K} &
\multicolumn{1}{c}{K/J} &
\colhead{Direct} &
\colhead{Ind 1} &
\colhead{Ind 2} \\
\colhead{(1)} &
\colhead{(2)} &
\colhead{(3)} &
\colhead{(4)} &
\colhead{(5)} &
\colhead{(6)} &
\colhead{(7)} &
\colhead{(8)} &
\colhead{(9)} &
\colhead{(10)} &
\colhead{(11)} \\
}
\startdata
DE 0255-4700 & 0.695 (L8.0/--) & 0.718 (L8.0/T0.5) & 0.684 (L8.0/--) & 1.007 (--/--) & 0.848 (T0.0/T0.0) & 0.686 & 0.603 & {\bf L9.0} & L8.5: & \nodata  \\
SD 1254-0122 & 0.501 (T2.0/T2.0) & 0.582 (T1.5/T2.5) & 0.494 (T2.0/T2.0) & 0.880 (T2.0/T2.0) & 0.576 (T2.0/T2.0) & 0.567 & 0.327 & {\bf T2.0} & T2.0 & T2.0  \\
2M 0559-1404 & 0.342 (T4.5/T4.0) & 0.454 (T4.5/T4.5) & 0.388 (T4.5/T4.0) & 0.472 (T4.5/T4.5) & 0.239 (T4.5/T4.5) & 0.468 & 0.192 & {\bf T4.5} & T4.5 & T4.5  \\
2M 1503+2525 & 0.239 (T5.0/T5.0) & 0.387 (T5.0/T5.0) & 0.332 (T5.0/T5.5) & 0.381 (T5.0/T5.5) & 0.200 (T5.0/T5.0) & 0.436 & 0.151 & {\bf T5.0} & T5.0 & T5.0  \\
2M 1901+4718 & 0.305 (T4.5/T4.5) & 0.444 (T4.5/T4.5) & 0.354 (T4.5/T5.0) & 0.437 (T4.5/T5.0) & 0.261 (T4.5/T4.5) & 0.504 & 0.172 & {\bf T5.0} & T4.5 & T4.5  \\
2M 2331-4718 & 0.235 (T5.0/T5.0) & 0.377 (T5.0/T5.5) & 0.332 (T5.0/T5.5) & 0.424 (T4.5/T5.0) & 0.225 (T5.0/T4.5) & 0.522 & 0.208 & {\bf T5.0} & T5.0 & T5.0  \\
2M 1231+0847 & 0.194 (T5.5/T5.5) & 0.353 (T5.5/T5.5) & 0.278 (T6.5/T6.0) & 0.347 (T5.5/T5.5) & 0.161 (T6.0/T6.0) & 0.401 & 0.156 & {\bf T5.5} & T6.0 & T5.5  \\
2M 1828-4849 & 0.175 (T6.0/T5.5) & 0.347 (T5.5/T5.5) & 0.310 (T6.0/T5.5) & 0.360 (T5.5/T5.5) & 0.197 (T5.0/T5.0) & 0.515 & 0.166 & {\bf T5.5} & T5.5 & T5.5  \\
\enddata
\tablecomments{Index subtypes in parentheses determined by direct comparison to spectral standard
ratios (Table~\ref{tab_indstandard}) or spectral ratio ranges (Table~\ref{tab_indrng}); see $\S$~4.2.2.
Uncertain classifications are indicated by ``:''; peculiar sources are indicated by ``p''.}
\end{deluxetable}

\begin{deluxetable}{llllllcc|lll}
\tabletypesize{\scriptsize}
\tablecaption{Spectral Indices and Classifications for NIRSPEC Spectra. \label{tab_nircspecclass}}
\tablewidth{0pt}
\rotate
\tablehead{
\colhead{} &
\multicolumn{7}{c}{} &
\multicolumn{3}{c}{Derived NIR SpT} \\
\cline{9-11}
\colhead{Name} &
\multicolumn{1}{c}{H$_2$O-J} &
\multicolumn{1}{c}{CH$_4$-J} &
\multicolumn{1}{c}{H$_2$O-H} &
\multicolumn{1}{c}{CH$_4$-H} &
\multicolumn{1}{c}{CH$_4$-K} &
\multicolumn{1}{c}{H$_2$O-K} &
\multicolumn{1}{c}{K/J} &
\colhead{Direct} &
\colhead{Ind 1} &
\colhead{Ind 2} \\
\colhead{(1)} &
\colhead{(2)} &
\colhead{(3)} &
\colhead{(4)} &
\colhead{(5)} &
\colhead{(6)} &
\colhead{(7)} &
\colhead{(8)} &
\colhead{(9)} &
\colhead{(10)} &
\colhead{(11)} \\
}
\startdata
2M 1632+1904 & 0.649 (T1.0/--) & 0.723 (L8.0/T0.5) & 0.696 (L8.0/--) & 1.042 (--/--) & 0.892 (L8.0/--) & 0.683 & 0.898 & {\bf L8.0} & L8.5: & \nodata  \\
2M 0310+1648 & \nodata & 0.679 (T1.0/T1.5) & \nodata & \nodata & \nodata & \nodata & \nodata & {\bf L9.0} & T1.0: & T1.5:  \\
Gl 337CD & 0.632 (T0.0/--) & 0.712 (L8.0/T1.0) & 0.697 (L8.0/--) & 0.973 (T0.0/T0.5) & 0.789 (L9.0/T0.0) & 0.685 & 0.712 & {\bf T0.0} & L9.0: & T0.5  \\
SD 0423-0414 & 0.605 (T0.0/--) & 0.656 (T1.0/T1.5) & 0.615 (L9.5/T0.0) & 0.956 (T0.0/T1.0) & 0.826 (T0.0/T0.0) & 0.658 & 0.558 & {\bf T0.0} & T0.0 & T0.5  \\
SD 0151+1244 & \nodata & 0.655 (T1.0/T1.5) & \nodata & \nodata & \nodata & \nodata & \nodata & {\bf T1.0} & T1.0: & T1.5:  \\
SD 0837-0000 & \nodata & 0.641 (T0.0/T2.0) & 0.569 (T1.0/T1.0) & 0.991 (L8.0/T0.0) & 0.710 (T1.0/T1.0) & 0.599 & 0.306 & {\bf T1.0} & T0.0: & T1.0  \\
SD 1254-0122 & 0.465 (T2.5/T2.5) & 0.554 (T2.0/T3.0) & 0.474 (T2.5/T2.5) & 0.886 (T2.0/T2.0) & 0.573 (T2.0/T2.0) & 0.548 & 0.294 & {\bf T2.0} & T2.0 & T2.5  \\
SD 1021-0304 & \nodata & 0.503 (T3.5/T4.0) & \nodata & \nodata & \nodata & \nodata & \nodata & {\bf T3.0} & T3.5: & T4.0:  \\
\enddata
\tablecomments{Index subtypes in parentheses determined by direct comparison to spectral standard
ratios (Table~\ref{tab_indstandard}) or spectral ratio ranges (Table~\ref{tab_indrng}); see $\S$~4.2.2.
Uncertain classifications are indicated by ``:''; peculiar sources are indicated by ``p''.}
\end{deluxetable}

\begin{deluxetable}{llllllcc|lll}
\tabletypesize{\scriptsize}
\tablecaption{Spectral Indices and Classifications for OSIRIS Spectra. \label{tab_osirisclass}}
\tablewidth{0pt}
\rotate
\tablehead{
\colhead{} &
\multicolumn{7}{c}{} &
\multicolumn{3}{c}{Derived NIR SpT} \\
\cline{9-11}
\colhead{Name} &
\multicolumn{1}{c}{H$_2$O-J} &
\multicolumn{1}{c}{CH$_4$-J} &
\multicolumn{1}{c}{H$_2$O-H} &
\multicolumn{1}{c}{CH$_4$-H} &
\multicolumn{1}{c}{CH$_4$-K} &
\multicolumn{1}{c}{H$_2$O-K} &
\multicolumn{1}{c}{K/J} &
\colhead{Direct} &
\colhead{Ind 1} &
\colhead{Ind 2} \\
\colhead{(1)} &
\colhead{(2)} &
\colhead{(3)} &
\colhead{(4)} &
\colhead{(5)} &
\colhead{(6)} &
\colhead{(7)} &
\colhead{(8)} &
\colhead{(9)} &
\colhead{(10)} &
\colhead{(11)} \\
}
\startdata
SD 0423-0414 & \nodata & 0.628 (T0.0/T2.0) & 0.615 (L9.5/T0.0) & 0.923 (T1.0/T1.5) & 0.822 (T0.0/T0.0) & 0.687 & 0.470 & {\bf T0.0} & T0.0 & T1.0:  \\
SD 1254-0122 & \nodata & 0.554 (T2.0/T3.0) & 0.483 (T2.5/T2.0) & 0.855 (T2.0/T2.0) & 0.540 (T2.5/T2.5) & 0.507 & 0.412 & {\bf T2.0} & T2.5 & T2.5  \\
2M 2254+3123 & \nodata & 0.439 (T4.5/T4.5) & 0.372 (T4.5/T4.5) & 0.520 (T4.0/T4.0) & \nodata & 0.386 & 0.204 & {\bf T4.0} & T4.5 & T4.5  \\
2M 0559-1404 & \nodata & 0.371 (T5.0/T5.5) & 0.320 (T6.0/T5.5) & 0.402 (T5.0/T5.0) & 0.216 (T5.0/T5.0) & 0.558 & 0.215 & {\bf T4.5} & T5.5 & T5.5  \\
2M 1534-2952 & \nodata & 0.364 (T5.5/T5.5) & 0.341 (T5.0/T5.0) & 0.397 (T5.0/T5.0) & 0.195 (T5.0/T5.0) & 0.452 & 0.145 & {\bf T5.0} & T5.0 & T5.0  \\
2M 0243-2453 & \nodata & 0.500 (T3.5/T4.0) & 0.203 (T7.5/T7.5) & 0.247 (T6.5/T6.5) & \nodata & 0.549 & 0.231 & {\bf T6.0:} & T6.0: & T6.0:  \\
2M 0937+2931 & \nodata & 0.311 (T6.0/T6.0) & 0.268 (T6.5/T6.5) & 0.261 (T6.5/T6.5) & \nodata & \nodata & 0.089 & {\bf T6.0p} & T6.5 & T6.5  \\
2M 1225-2739 & \nodata & 0.306 (T6.0/T6.0) & 0.259 (T6.5/T6.5) & 0.278 (T6.5/T6.0) & \nodata & 0.298 & 0.137 & {\bf T6.0} & T6.5 & T6.0  \\
\enddata
\tablecomments{Index subtypes in parentheses determined by direct comparison to spectral standard
ratios (Table~\ref{tab_indstandard}) or spectral ratio ranges (Table~\ref{tab_indrng}); see $\S$~4.2.2.
Uncertain classifications are indicated by ``:''; peculiar sources are indicated by ``p''.}
\end{deluxetable}

\begin{deluxetable}{llllllcc|lll|l}
\tabletypesize{\scriptsize}
\tablecaption{Spectral Indices and Classifications for Other Published Spectra. \label{tab_litclass}}
\tablewidth{0pt}
\rotate
\tablehead{
\colhead{} &
\multicolumn{7}{c}{} &
\multicolumn{3}{c}{Derived NIR SpT} & \\
\cline{9-11}
\colhead{Name} &
\multicolumn{1}{c}{H$_2$O-J} &
\multicolumn{1}{c}{CH$_4$-J} &
\multicolumn{1}{c}{H$_2$O-H} &
\multicolumn{1}{c}{CH$_4$-H} &
\multicolumn{1}{c}{CH$_4$-K} &
\multicolumn{1}{c}{H$_2$O-K} &
\multicolumn{1}{c}{K/J} &
\colhead{Direct} &
\colhead{Ind 1} &
\colhead{Ind 2} &
\colhead{Ref} \\
\colhead{(1)} &
\colhead{(2)} &
\colhead{(3)} &
\colhead{(4)} &
\colhead{(5)} &
\colhead{(6)} &
\colhead{(7)} &
\colhead{(8)} &
\colhead{(9)} &
\colhead{(10)} &
\colhead{(11)} &
\colhead{(12)} \\
}
\startdata
2M 0920+3517 & \nodata & 0.739 (L9.0/T0.5) & \nodata & 0.867 (T2.0/T2.0) & 0.901 (L8.0/--) & 0.657 & 0.610 & {\bf T0.0p} & L9.5: & T1.5: & 1 \\
$\epsilon$ Indi Ba & \nodata & \nodata & 0.593 (T0.5/T0.5) & 0.904 (T1.5/T1.5) & \nodata & \nodata & \nodata & {\bf T1.0} & T1.0 & T1.0 & 2 \\
SD 1254-0122 & 0.491 (T2.0/T2.0) & 0.560 (T2.0/T3.0) & 0.507 (T2.0/T2.0) & 0.846 (T2.0/T2.0) & 0.564 (T2.0/T2.5) & 0.549 & 0.354 & {\bf T2.0} & T2.0 & T2.5 & 3 \\
IFA 0230-Z1 & \nodata & \nodata & \nodata & 0.646 (T3.5/T3.5) & \nodata & \nodata & \nodata & {\bf T3.0:} & T3.5: & T3.5: & 4 \\
SD 1750+1759 & 0.471 (T2.5/T2.5) & 0.507 (T3.5/T4.0) & 0.475 (T2.5/T2.5) & 0.605 (T3.5/T3.5) & 0.416 (T3.5/T3.0) & 0.602 & 0.176 & {\bf T3.5} & T3.0 & T3.0 & 3 \\
2M 0559-1404 & 0.333 (T4.5/T4.5) & 0.451 (T4.5/T4.5) & 0.426 (T4.0/T3.5) & 0.490 (T4.5/T4.5) & 0.256 (T4.5/T4.5) & 0.627 & 0.173 & {\bf T4.5} & T4.5 & T4.5 & 5 \\
$\epsilon$ Indi Bb & \nodata & \nodata & 0.345 (T5.0/T5.0) & 0.306 (T6.0/T6.0) & \nodata & \nodata & \nodata & {\bf T6.0} & T5.5 & T5.5  & 2 \\
NT 1205-0744 & 0.112 (T6.5/T7.0) & 0.326 (T6.0/T6.0) & 0.140 (T8.0/T8.0) & 0.244 (T6.5/T6.5) & 0.240 (T4.5/T4.5) & 0.376 & 0.159 & {\bf T7.0} & T6.5: & T6.5: & 6 \\
\enddata
\tablecomments{Index subtypes in parentheses determined by direct comparison to spectral standard
ratios (Table~\ref{tab_indstandard}) or spectral ratio ranges (Table~\ref{tab_indrng}); see $\S$~4.2.2.
Uncertain classifications are indicated by ``:''; peculiar sources are indicated by ``p''.}
\tablerefs{(1) \citet{nak01}; (2) \citet{mcc04}; (3) \citet{nak04}; (4) \citet{liu02}; (5) \citet{me00c}; (6) \citet{cub99}}
\end{deluxetable}

\clearpage

\begin{deluxetable}{llll}
\tabletypesize{\scriptsize}
\tablecaption{Comparison of T Dwarf Spectral Types \label{tab_tcomp}}
\tablewidth{0pt}
\tablehead{
 & \multicolumn{3}{c}{NIR Spectral Type} \\
\cline{2-4}
\colhead{Object} &
\colhead{\bf New} &
\colhead{B02} &
\colhead{G02} \\
\colhead{(1)} &
\colhead{(2)} &
\colhead{(3)} &
\colhead{(4)} \\
}
\startdata
SDSS 2047-0718 & {\bf T0.0:} &  \nodata &  L9.5$\pm$1 \\
SDSS 1104+5548 & {\bf T0.0:} & \nodata  & L9.5$\pm$1.5  \\
SDSS 1207+0244 & {\bf T0.0} & \nodata  &  T0 \\
SDSS 1516+0259 & {\bf T0.0:} & \nodata  & T0$\pm$1.5  \\
SDSS 0423-0414 & {\bf T0.0} & T0\tablenotemark{a} & T0  \\
SDSS 0151+1244 & {\bf T0.5} & T0.5  & T1$\pm$1  \\
SDSS 0837-0000 & {\bf T1.0} &  T1 &  T0.5 \\
$\epsilon$ Indi Ba & {\bf T1.0} & T1  & T0.5  \\
\enddata
\tablenotetext{a}{Based on revised indices similar to Table~\ref{tab_indices} \citep{me04a,tin05}.}
\end{deluxetable}

\clearpage

\begin{deluxetable}{llccccccccccl}
\tabletypesize{\tiny}
\setlength{\tabcolsep}{0.02in}
\rotate
\tablecaption{T Dwarf Compendium \label{tab_allt}}
\tablewidth{0pt}
\tablehead{
 & \colhead{NIR} & \multicolumn{3}{c}{J2000 Coordinates\tablenotemark{a}} &
\multicolumn{1}{c}{SDSS\tablenotemark{b}} &
 \multicolumn{3}{c}{2MASS\tablenotemark{c}} \\
\cline{3-5} \cline{7-9}
\colhead{Name} &
\colhead{SpT\tablenotemark{d}} &
\colhead{$\alpha$} &
\colhead{$\delta$} &
\colhead{Epoch} &
\colhead{$z$} &
\colhead{$J$} &
\colhead{$H$} &
\colhead{$K_s$} &
\colhead{$\pi$} &
\colhead{$\mu$} &
\colhead{$\phi$} &
\colhead{Ref\tablenotemark{e}} \\
 & & & & &
\colhead{(AB)} &
\colhead{(mag)} &
\colhead{(mag)} &
\colhead{(mag)} &
\colhead{(marcs)} &
\colhead{($\arcsec$ yr$^{-1}$)} &
\colhead{($\degr$)} \\
\colhead{(1)} &
\colhead{(2)} &
\colhead{(3)} &
\colhead{(4)} &
\colhead{(5)} &
\colhead{(6)} &
\colhead{(7)} &
\colhead{(8)} &
\colhead{(9)} &
\colhead{(10)} &
\colhead{(11)} &
\colhead{(12)} &
\colhead{(13)} \\
}
\startdata
SDSS J000013.54+255418.6 & T4.5 & 00$^h$00$^m$13$\fs$54 & +25$\degr$54$\arcmin$18$\farcs$0 & 1998.76 & 18.48$\pm$0.04 & 15.06$\pm$0.04 & 14.73$\pm$0.07 & 14.84$\pm$0.12 & \nodata & \nodata & \nodata & {\bf 1} \\
2MASS J00345157+0523050 & T6.5 & 00$^h$34$^m$51$\fs$57 & +05$\degr$23$\arcmin$05$\farcs$0 & 2000.67 & \nodata & 15.54$\pm$0.05 & 15.44$\pm$0.08 & $>$ 16.2 & \nodata & 0.68$\pm$0.06 &  72 & {\bf 2} \\
2MASS J00501994$-$3322402 & T7 & 00$^h$50$^m$19$\fs$94 & $-$33$\degr$22$\arcmin$40$\farcs$2 & 1998.87 & \nodata & 15.93$\pm$0.07 & 15.84$\pm$0.19 & 15.24$\pm$0.19 & \nodata & 1.5$\pm$0.1 & 307 & {\bf 3} \\
SDSS J015141.69+124429.6 & T0.5 & 01$^h$51$^m$41$\fs$55 & +12$\degr$44$\arcmin$30$\farcs$0 & 1997.70 & 19.40$\pm$0.07 & 16.57$\pm$0.13 & 15.60$\pm$0.11 & 15.18$\pm$0.19 & 47$\pm$3 & 0.743$\pm$0.004 &  93 & {\bf 4},24 \\
SDSS J020742.48+000056.2 & T4.5 & 02$^h$07$^m$42$\fs$84 & +00$\degr$00$\arcmin$56$\farcs$4 & 2000.63 & 20.08$\pm$0.12 & 16.63$\pm$0.05\tablenotemark{f} & 16.66$\pm$0.05\tablenotemark{f} & 16.62$\pm$0.05\tablenotemark{f} & 35$\pm$10 & 0.156$\pm$0.011 &  96 & {\bf 4},24,25 \\
IfA 0230$-$Z1 & T3 & 02$^h$26$^m$37$\fs$60 & +00$\degr$51$\arcmin$54$\farcs$7 & 2000.81 &  & 18.17$\pm$0.03\tablenotemark{g} & 17.83$\pm$0.04\tablenotemark{g} & \nodata & \nodata & \nodata & \nodata & {\bf 5} \\
2MASS J02431371$-$2453298 & T6 & 02$^h$43$^m$13$\fs$71 & $-$24$\degr$53$\arcmin$29$\farcs$8 & 1998.86 & \nodata & 15.38$\pm$0.05 & 15.14$\pm$0.11 & 15.22$\pm$0.17 & 94$\pm$4 & 0.355$\pm$0.004 & 234 & {\bf 6},24 \\
2MASS J03480772$-$6022270 & T7 & 03$^h$48$^m$07$\fs$72 & $-$60$\degr$22$\arcmin$27$\farcs$0 & 1999.88 & \nodata & 15.32$\pm$0.05 & 15.56$\pm$0.14 & 15.60$\pm$0.23 & \nodata & 0.77$\pm$0.04 & 201 & {\bf 7} \\
\enddata
\tablenotetext{a}{If detected in 2MASS, coordinates are from the
2MASS All Sky Point Source Catalog \citep{cut03}, epoch $\sim$1997-2001;
otherwise coordinates are from discovery reference.}
\tablenotetext{b}{SDSS $z$ AB magnitudes from SDSS Data Release 4 \citep{ade05} or the literature
\citep{leg00,geb02,kna04}.}
\tablenotetext{c}{Unless otherwise noted.}
\tablenotetext{d}{Near infrared spectral type on unified scheme; see $\S$~4.}
\tablenotetext{e}{Discovery reference in boldface type, followed by references for additional photometric
and astrometric data.}
\tablenotetext{f}{MKO $JHK$ from \citet{leg02b} or \citet{kna04}.}
\tablenotetext{g}{MKO $JH$ photometry from \citet{liu02}.}
\tablenotetext{h}{$JHK$ photometry from \citet{zap02}.}
\tablenotetext{i}{Coordinates for Gliese 229B determined from 2MASS coordinates of Gliese 229A
and offsets from \citet{gol98}.}
\tablenotetext{j}{UKIRT $JHK$ photometry from \citet{leg99}.}
\tablenotetext{k}{$JK$ photometry from \citet{cub99}.}
\tablenotetext{l}{Spectral types from resolved spectroscopy (see Table~\ref{tab_litclass}).}
\tablerefs{(1) \citet{kna04}; (2) \citet{me04a}; (3) \citet{tin05}; (4) G02;
(5) \citet{liu02}; (6) B02; (7) \citet{me03e};
(8) \citet{cru04}; (9) \citet{zap02}; (10) \citet{me00c};
(11) \citet{nak95}; (12) \citet{leg00}; (13) \citet{wil01}; (14) \citet{kir00};
(15) \citet{me99}; (16) \citet{cub99}; (17) \citet{tsv00}; (18) \citet{me00a}; (19) \citet{me03a};
(20) \citet{str99}; (21) Cruz et al.\ in prep.; (22) \citet{ell05}; (23) \citet{sch03};
(24) \citet{vrb04}; (25) \citet{leg02b}; (26) HIPPARCOS \citep{esa97};
(27) \citet{me05c};
(28) \citet{mrt03}; (29) \citet{dah02}; (30) \citet{opp95}; (31) \citet{leg99};
(32) \citet{mcl03}; (33) \citet{me05b}; (34) \citet{nak01}; (35) \citet{rei01};
(36) Burgasser et al., in prep.; (37) \citet{tin03}; (38) \citet{me03b}; (39) \citet{mcc04}
}
\end{deluxetable}

\clearpage

\clearpage

\end{document}